\documentclass[fleqn,usenatbib]{mnras}

\usepackage{newtxtext,newtxmath}
\usepackage[T1]{fontenc}
\usepackage{ae,aecompl}
\usepackage{etoolbox}

\usepackage{graphicx}	% Including figure files
\usepackage{amsmath}	% Advanced maths commands
\usepackage{multirow}
\usepackage{epstopdf}
\usepackage[figure,figure*]{hypcap}
\usepackage{float}
\usepackage{ae,aecompl}
\usepackage{etoolbox}

\newcommand{\comm}[1]{}
\defcitealias{2019MNRAS.490.1539R}{R19}

\DeclareMathOperator{\sech}{sech}
\title[The problem of dust attenuation]{The problem of dust attenuation in photometric decomposition of edge-on galaxies and possible solutions}
\author[S. S. Savchenko et al.]{
Sergey S. Savchenko,$^{1,2,3}$\thanks{E-mail: s.s.savchenko@spbu.ru}
Denis M. Poliakov,$^{1}$
Aleksandr V. Mosenkov,$^{4}$
Anton A. Smirnov,$^{1}$
\newauthor
Alexander A. Marchuk,$^{1,2}$
Vladimir B. Il'in,$^{1,2, 5}$
George A. Gontcharov,$^{1}$
Jonah Seguine$^{4}$,
and M. Baes$^{6}$
\\
$^{1}$Central Astronomical Observatory at Pulkovo of RAS, Pulkovskoye Chaussee 65/1, 196140 St. Petersburg, Russia\\
$^{2}$St.Petersburg State University, 7/9 Universitetskaya nab., St.Petersburg, 199034, Russia\\
$^{3}$Special Astrophysical Observatory, Russian Academy of Sciences, 369167 Nizhnij Arkhyz, Russia\\
$^{4}$Department of Physics and Astronomy, N283 ESC, Brigham Young University, Provo, UT 84602, USA\\
$^{5}$Saint Petersburg University of Aerospace Instrumentation, Bol. Morskaya ul. 67A, St. Petersburg 190000, Russia\\
$^{6}$Sterrenkundig Observatorium, Universiteit Gent, Krijgslaan 281 S9, 9000 Gent, Belgium\\
}

\date{Accepted XXX. Received YYY; in original form ZZZ}

\pubyear{2023}

\begin{document}
\label{firstpage}
\pagerange{\pageref{firstpage}--\pageref{lastpage}}
\maketitle

\begin{abstract}
  The presence of dust in spiral galaxies affects the ability of photometric
  decompositions to retrieve the parameters of their main structural components. For galaxies in an edge-on orientation, the optical
  depth integrated over the line-of-sight is significantly higher than for those with intermediate or face-on inclinations, so it is only natural to expect that for edge-on galaxies, dust attenuation should severely influence measured structural parameters. In this paper, we use
  radiative transfer simulations to generate a set of synthetic images of edge-on galaxies which are then analysed via
  decomposition. Our results demonstrate that for edge-on galaxies, the observed systematic errors of the fit parameters are
  significantly higher than for moderately inclined galaxies. Even for models with a relatively low dust content, all structural
  parameters suffer offsets that are far from negligible. In our search for ways to reduce the impact of dust on retrieved structural parameters, we test
  several approaches, including various masking methods and an analytical model that incorporates dust absorption. We
  show that using such techniques greatly improves the reliability of decompositions for edge-on galaxies.
\end{abstract}

\begin{keywords}
Galaxy: structure - fundamental parameters - formation - disc - bulge
\end{keywords}

%%%%%%%%%%%%%%%%%%%%%%%%%%%%%%%%%%%%%%%%%%%%%%%%%%

%%%%%%%%%%%%%%%%% BODY OF PAPER %%%%%%%%%%%%%%%%%%

\section{Introduction}
\label{sec:intro}
Measuring the physical properties of galaxies is one of the cornerstones of extragalactic astrophysics, because all
theories of galaxy formation and evolution should be supported by observational data. 
Disc galaxies which are visible in an edge-on orientation (i.e. inclined at $i\approx90\degr$) are of special interest in this regard because they are the only targets that facilitate a direct study of the vertical structures of disc galaxies. For example, the vertical distributions of stars, gas, and dust, as well as the possible presence of different sub-components (such as thin and thick discs), which properties are often described via various galaxy scaling relations (such as the dependence of the disc flattening on the relative mass of a spherical component, including a dark matter halo) can only be explored in edge-on galaxies \citep[see e.g.][]{Kylafis1987, Xilouris1999,2010MNRAS.401..559M,Bizyaev2014,2018A&A...610A...5C,2022MNRAS.515.5698M}. 

This utility of edge-on galaxies is easily recognized due to the existence of special catalogues which were created specifically for studying the three-dimensional structure of disc galaxies. For example, the RFGC catalogue \citep{Karachentsev1999} contains 4236 thin edge-on spiral galaxies over the whole sky. The EGIS catalog \citep{Bizyaev2014} provides structural parameters of stellar discs (the disc scale length and scale height, as wells as the disc central surface brightness) for almost 6000 galaxies using the Sloan Digital Sky Survey (SDSS, \citealt{2000AJ....120.1579Y}) observations in several optical wavebands. The EGIPS catalogue \citep{Makarov2022} contains 16551 edge-on galaxies from the Pan-STARRS survey \citep{Chambers2016, Flewelling2020}.

The most widely used approach to acquire the structural parameters of galaxies is performing a photometric decomposition of their images. The main idea behind this process is to adopt an analytical model to describe the
observed surface brightness distribution in a galaxy image and find the optimal parameters for such a model that yield the fewest discrepancies with the real image. 
There exist a number of software packages (see \citealt{2002AJ....124..266P, Vika2013, DeGeyter2013, Erwin2015}) which were specifically
designed to perform photometric decompositions of galaxies (e.g. \citealt{Gadotti2009, Lackner2012, Bottrell2019}, and many others). For example, in almost two thousand refereed publications to date, the GALFIT code has been used to retrieve the structural parameters of galaxies with various morphologies, at different wavelengths, and in a wide range of redshifts.

Although at a first glance the main idea behind the decomposition process looks rather straightforward, there are
various obstacles that must be overcome on the way to a solid and robust estimation of galaxy parameters. For example, even the
model selection can be a problem, especially when working with a large sample of objects \citep{Lingard2020}.
Another complicating factor is image smearing caused by a point spread function (PSF) due to atmospheric turbulence (seeing) and the physical diffraction limit. The general rule is that the smaller the galaxy component, the larger the influence of the PSF on the
retrieved structural parameters \citep{Trujillo2001, Trujillo2001b, Gadotti2009}, and this is especially true for edge-on galaxies \citep{2014A&A...567A..97S,2015A&A...577A.106S}.

In this article, we focus on another important issue for galaxy photometric decompositions that manifests particularly strongly for edge-on galaxies, dust attenuation. The dust
distributed in a galaxy absorbs, scatters, and re-emits its stellar light, resulting in an observed surface brightness distribution for a
galaxy image, coupled with a mass-to-stellar luminosity ratio as a function of both radius and wavelength, that does not reflect the actual mass surface density distribution over the galaxy body. This suggests that the measured structural parameters can be affected in a manner that is challenging to predict.

One possible solution to this problem is to perform radiative transfer modelling that includes the interaction between the photons and dust. The complexity of such approaches have grown over time. For example, \citet{Disney1989}
provided an analysis of several simple geometric models including a ``slab'' model where the galaxy is considered to
be a flat disc with a uniform mixture of stars, dust, and gas; a ``screen'' model with a stellar disc covered by a dust
absorbing screen lying above the stars; and a ``sandwich'' model where a thin uniform dust disc is located inside
of a relatively thicker stellar disc. \citet{Byun1994} performed numerical radiative transfer modelling of a
three-dimensional galaxy model with various dust contents visible at different viewing angles to study how dust
attenuation changes the main observables of the galaxy (ellipticity, surface brightness, exponential scale, etc.). The dust's
impact on attenuation in a galaxy as a function of wavelength was studied by \citet{Ferrara1999} for
a set of different galaxy models (mimicking spiral and elliptical galaxies) and by \citet{Tuffs2004} who considered exponential discs and de Vaucouleurs bulges as separate components. A combined bulge+disc model was studied in \citet{Pierini2004}. 

Nowadays, there are various tools that allow one to carry out radiative transfer modeling for complex multicomponent galaxies with dust. For example, \citet{Popescu2000} describe such an approach and its application to an edge-on galaxy NGC~891. Other examples of radiative transfer programs include the {\small TRADING} code \citep{Bianchi2008}, the {\small DART-RAY} code \citep{Natale2014, Natale2017}, and the {\small FITSKIRT} software \citep{DeGeyter2013}. For example, using {\small FITSKIRT}, it is possible to fit a galaxy image with a predefined model consisting of multiple stellar and dust components. A significant drawback of such an approach is its extreme computational cost \citep{Mosenkov2018}, and, thus, it is only useful when trying to model individual edge-on galaxies \citep{Xilouris1997, Xilouris1998, Popescu2000, Baes2010, Bianchi2011, DeLooze2012, 2012ApJ...746...70S, DeGeyter2013, Mosenkov2016} or small samples of galaxies \citep{Xilouris1999, Bianchi2007, DeGeyter2014, Mosenkov2018, Natale2022}.

Another approach frequently used to investigate the effect opacity has on measured structural parameters in disc galaxies is to use radiative transfer simulations to create a mock galaxy image. This involves using a given, a-priori known model and then decomposing the image. In this case, one can explore how exactly the presence of dust affects the galactic parameters measured by using a regular decomposition technique, that is, without including the radiative transfer or any other dust compensation method. This was done by \citet{Gadotti2010}, who investigated the behaviour of a couple of models of disc galaxies for a range of dust optical depth values and for inclination angles ranging from 15 to 60 degrees. A similar
approach was adopted by \citet{Pastrav2013a} and \citet{Pastrav2013b} where a set of corrections for the measured decomposition parameters were computed. These corrections were applied in \citet{Pastrav2020} to find the intrinsic (i.e. dust-corrected)
parameters for several real galaxies.

In this paper, we concentrate on the effects of dust attenuation on the parameters of edge-on galaxies measured via a standard decomposition analysis. By building our article on the work done by the aforementioned studies, we opt to go further in this analysis and add some new features, such as
\begin{itemize}
\item using three-dimensional decomposition models with line-of-sight integration instead of traditional two-dimensional fitting, which will allow us to treat the structure of edge-on galaxies to the fullest;
\item simulating real observations by accounting for instrument PSFs, transmission curves, and noise parameters;
\item running simulations for a set of models to explore how mid-plane dust lanes impact galaxy structural parameters.
\end{itemize}

The other important goal of this study is to investigate various techniques to compensate for the presence of dust during the decomposition
process (aside from the time-consuming radiative transfer approach). Is it possible to modify the decomposition procedure to make the derived parameters more reliable without a significant increase to computational time? If so, can we
apply this approach to a large sample of edge-on galaxies? The answers to these questions are of high importance for the ongoing work with the EGIPS catalogue \citep{Makarov2022} where we aim to perform a mass decomposition of edge-on galaxies with three-dimensional models using ``dust contaminated'' optical observations.

The rest of the article is organized as follows. In Section~\ref{sec:algorithms}, we describe our algorithms for
synthetic image creation and decomposition with and without correcting for dust. In Section~\ref{sec:simulations}, we demonstrate the results of our simulations including the dust impact on the derived decomposition parameters, and the results of applying different techniques to compensating for the presence of the dust. In Section~\ref{sec:real_galaxies}, we employ the decomposition methods for retrieving the structural parameters for a couple of real galaxies, with taking a dust component into account. We state our conclusions in Section~\ref{sec:conclusions}. Appendix~\ref{sec:appendix_neural} contains some technical details about the training of the neural network used throughout the paper.

\section{The algorithm}
\label{sec:algorithms}
In this section, we describe in detail our algorithms to investigate the dust impact on the decomposition results
and propose several ways to ease these dust effects. The overall pipeline looks as follows. For a set of input parameters we create a three-dimensional model
of a galaxy, and transform it into a FITS-file  by projecting it on an image plane. To mimic real observations, we include a smearing effect by a PSF and add read and photon noise. We then run a standard decomposition technique
to obtain the observed structural parameters of a galaxy in order to compare them with the input ones. After that, we
run a series of fits using various methods, by accounting for the dust presence and without accounting for it, to see how these can amend the observed parameters.

\subsection{Model functions and their parameters}
\label{sec:input_params}
In this work we consider a three-component model of a galaxy with a stellar disc, bulge, and a dust disc. As a disc model, we
adopt a three-dimensional isothermal disc that follows an exponential luminosity density profile in the radial direction and
a $\sech^2$ law perpendicular to the galaxy plane \citep{vanderKruit1981}:

\begin{equation}
  \label{eq:expdisc}
  \rho_{\mathrm{disc}}(r, z) = \rho_{0, \mathrm{disc}} \exp \left( - \frac{r}{h_{\mathrm{disc}}} \right) \sech^2 \left( \frac{z}{z_{\mathrm{disc}}} \right).
\end{equation}

The disc model has two geometric parameters: its radial exponential scale $h_{\mathrm{disc}}$ and the vertical scale
$z_{\mathrm{disc}}$. The third parameter is the central luminosity density $\rho_{0, \mathrm{disc}}$, which governs the luminosity of the disc, but for our purposes it is more convenient to work directly with the disc's total luminosity: 

\begin{equation}
L_{\mathrm{disc}} = 4\pi \rho_{0, \mathrm{disc}} h_{\mathrm{disc}}^2 z_{\mathrm{disc}}.
\end{equation}

Even though real galaxies can demonstrate more complex disc structures (such as the existence of two embedded stellar discs), we do not include such complexity in our simulations, because it can only be studied for the closest galaxies with better spatial resolution, and most decomposition studies utilise a single disc model.

To model a central component, we use the well-known 
S\'{e}rsic function \citep{Sersic1963, Sersic1968} that is often used to describe galactic bulges, and which have the following projected surface brightness profile:
\begin{equation}
  \label{eq:sersic}
  I(r) = I_0 \exp \left[ -\nu_n \left( \frac{r}{r_e} \right)^{\frac{1}{n}} \right].
\end{equation}
The corresponding three-dimensional density distribution that is required for our work can be found through an Abel inversion:
\begin{equation}
  \label{eq:abel}
  \rho_{\mathrm{bulge}}(r) = - \frac{1}{\pi}\int_r^{+\infty}\frac{dI}{dR} \frac{dR}{\sqrt{R^2 - r^2}}.
\end{equation}

In the literature there are various approaches on how to solve this integral analytically (with special functions) or
numerically \citep{Prugniel1997, Lima1999, Baes2011, Baes2011b, Vitral2020}. The bulge has three main geometric
parameters (the value $\nu_n$ in (\ref{eq:sersic}) is a normalisation constant): its effective radius $r_e$, a S\'{e}rsic parameter $n$, and a bulge oblateness $q_{\mathrm{bulge}}$, and a parameter, that governs the overall bulge brightness,
the central luminosity density $\rho_{0, \mathrm{bulge}}$, but as before, it is more convenient to use the total luminosity as a free parameter:
\begin{equation}
  L_{\mathrm{bulge}} = \rho_{0, \mathrm{bulge}} r_e^3  
\end{equation}

We describe the dust component with the same isothermal disc model as the stellar disc (\ref{eq:expdisc}) and it has the
following set of geometric parameters: $h_{\mathrm{dust}}$ and $z_{\mathrm{dust}}$. Following the previous works of
\cite{Gadotti2010, Pastrav2013b}, we parameterize the dust content not by its central density $\rho_{0, \mathrm{dust}}$,
but by using the central face-on optical depth $\tau$, that is, an integral characteristic of the galaxy opacity which can be computed by a line-of-sight integral drawn through the center of a face-on oriented model (\ref{eq:expdisc}):
\begin{equation}
    \label{eq:tau_def}
    \tau = \int_{-\infty}^\infty \kappa \rho_{\mathrm{dust}} (0, z)\, dz = 2\kappa z_{\mathrm{dust}} \cdot \rho_{0, dust},
\end{equation}
where $\kappa$ is the extinction coefficient that depends on the dust mixture and the observed wavelength. Throughout this paper, we measure $\tau$ values in the $V$ band to be consistent with \citet{Gadotti2010}, although different normalizations are used in the literature (for example, the $B$ band in \citealt{Pastrav2013a, Pastrav2013b}).

The galaxy as a whole has the following free parameters (apart from parameters specific for separate components): the total bolometric luminosity $L_{\mathrm{total}}$, the bulge-to-total luminosity ratio
$B/T$ (these two parameters defy the actual values of $L_{\mathrm{disc}}$ and $L_{\mathrm{bulge}}$),
the luminosity distance $D_\mathrm{L}$, and the inclination $i$.

Theoretically, a change in any parameter in a galaxy model can lead to changes in random and systematic errors in any other parameters, but it is difficult to make a
set of models that cover this parameter space well enough to study any possible inter-combinations between all
parameters. To achieve the goal of this study, we settle upon the following strategy to create the model grid. We
start with a single model which has the parameters of a typical disc galaxy (see, for example \citealt{Gadotti2009}), listed in Tab.~\ref{tab:basic_model}. Apart from the described set of geometric parameters above, this list also contains ages of stellar populations for the bulge and the disc such that the disc contains a younger stellar population than the bulge (4 Gyr versus 11 Gyr of a bulge), a galactic average metallicity, and a dust mixture that has mean properties found in \citet{Zubko2004}. Hereafter, we will call this model a \textit{basic model}. Then, we variate some parameters of this model leaving others fixed to see how these variations affect
the decomposition results.

\begin{table}
  \centering
  \begin{tabular}[h]{ll}
    \hline
    Parameter & Value \\
    \hline
    Bulge effective radius, $r_e$ & 900 pc\\
    Bulge S\'ersic index, $n$ & 4 \\
    Bulge oblateness,  $q$ & 0.0 \\
    Bulge stellar population age & 11 Gyr\\
    Disc radial scalelength, $h_{\mathrm{disc}}$ & 4000 pc\\
    Disc vertical scalelength, $z_{\mathrm{disc}}$ & 400 pc\\
    Disc stellar population age & 4 Gyr\\
    Dust radial scalelength, $h_{\mathrm{dust}}$ & 4000 pc\\
    Dust vertical scalelength, $z_{\mathrm{dust}}$ & 150 pc\\
    Dust mixture  & \cite{Zubko2004}\\
    Galaxy bolometric luminosity, $L_{\mathrm{tot}}$ & $10^{11} L_{\odot}$\\
    Galaxy average metallicity & 0.02 \\
    Bulge-to-total luminosity ratio, $B/T$ & 0.2\\
    \hline
  \end{tabular}
  \caption{The basic model parameters}
  \label{tab:basic_model}
\end{table}

\subsection{Synthetic images}
\label{sec:synthimag}
Creating a galaxy model with a dust component requires accurate Monte-Carlo radiative transfer
simulations. For this purpose, we use the state-of-art radiative transfer code {\small SKIRT} \citep{Baes2011c, Camps2015, Camps2020}.
{\small SKIRT} allows one to generate panchromatic simulations of a galaxy with provided parameters for the specified structural components (in our case, these are a
bulge, disc, and dust). The output data cube (a collection of two-dimensional images) contains different layers with snapshots of the galaxy model for the chosen set of wavelengths.

Next, each synthesized galaxy image from the data cube can be transformed into a new mock image to simulate observational effects which are always present in real observations. Here, we take into account specific instrument transmission curves 
by multiplying all $N$ individual layers of the model data cube $I_j(x, y)$ by the instrument response for the corresponding wavelength $f_j$. Then we coadd the layers into a single image $I(x, y)$, making sure to take into account the
wavelength width of each layer $W_j$:
\begin{equation}
  \label{eq:combining_images}
  I(x, y) = \sum_{j=1}^N I_j(x, y) f_j W_j.
\end{equation}
After that, it is necessary to convolve the obtained image with a PSF to simulate the effects of
atmospheric and telescopic blurring. Finally, we add Gaussian and Poisson noise to the image.

For this study we decided to use the SDSS $r$ waveband as the instrument system because
this survey is widely used as a data source for galaxy decompositions. Since this is an optical survey, dust attenuation can be high for edge-on galaxies. Therefore, we generate our mock galaxy images using the instrument and PSF parameters (filter transmission curve, full width at half maximum (FWHM) of PSF, gain, and readnoise values) specific for the SDSS $r$-band instrument. The average values of these characteristics (gain value equal to 4.75 electrons per count and dark variance equal to 1.32 electrons) are taken from the SDSS website\footnote{\url{https://dr12.sdss.org/datamodel/files/BOSS_PHOTOOBJ/frames/RERUN/RUN/CAMCOL/frame.html}}. An example of a model with a face-on optical depth $\tau=1.0$
is shown on the top panel of Fig.~\ref{fig:masking_examples}.

\subsection{Regular decomposition}
\label{sec:decomposition}
When a mock galaxy image is ready, we use the {\small IMFIT} code \citep{Erwin2015} to perform our decomposition of the image. One of the standard functions in {\small IMFIT} is a three-dimensional model of the disc which allows one to account for the projection effects and fit the galaxy inclination. Employing this function, we can reliably investigate the vertical structure of a highly inclined disc galaxy, whereas two-dimensional models of an edge-on disc work, strictly speaking, for perfect edge-on orientations only.

The {\small IMFIT} package allows one to take the PSF into account during fitting. We provide the same PSF image that we used to blur the mock image in Sec.~\ref{sec:synthimag}.

The {\small IMFIT} list of models includes both a S\'{e}rsic function and a three-dimensional isothermal disc, so the
output results can be directly compared with the input parameters from the {\small SKIRT} model. The only necessary step is to convert the output geometrical parameters of the {\small IMFIT}, which are given in pixels, back to parsecs using the model distance.

We emphasize that the model of the disc component that is used for the decomposition has an actual three-dimensional
volume brightness distribution. To produce a projected two-dimensional model image, {\small IMFIT} performs
integration along the line-of-sight $s$ of the volume luminosity density for each pixel $(x, y)$ of the image:
\begin{equation}
  \label{eq:los_integration}
  I(x, y) = \int_{-\infty}^{+\infty} \rho(s)\, \mathrm{d}s.
\end{equation}
This approach requires considerably more computational time than directly computing the two-dimensional exponential surface
brightness distribution. However, the advantage of this method is that it can give accurate results for models in an orientation {\it close} to edge-on. The insufficiency of
a simple two-dimensional model to describe an actual three-dimensional disc is clearly shown in \citet{Pastrav2013b}
where their decomposition results diverge quickly near the edge-on orientation. In \cite{Gadotti2010}, the highest inclination
considered was 60\degr which allowed them to elude this problem.

\subsection{Decomposition with a dust correction}
\label{sec:decomp_dust_corr}
As we will see in Sec.~\ref{sec:simulations}, the effects of a dust component on the derived parameters of edge-on
galaxies are enormous, so in some cases these make the fit results completely unreliable. To mitigate this problem,
we try a number of modifications on the regular decomposition technique. In this study, we test two approaches: (i) We use different masks for the dust lane to exclude ``dusty'' pixels from our decomposition and (ii) we modify the decomposition model to account for the dust attenuation. Below we describe these two methods in detail.

\subsubsection{Masking}
\label{sec:masking}
The vertical scale height of the dust component in galaxies is smaller than that of the stellar disc. As a result, dust attenuation in edge-on galaxies appears as a dark narrow lane along the mid-plane of
the galactic disc, whereas below and above this lane the galaxy appears less obscured. During decomposition, the
analytical model fits both attenuated and unattenuated regions of the galaxy which results in systematic errors
on the model's parameters. It therefore seems promising that, by masking out the attenuated dust lane in the galaxy image, the
fitting procedure can create a better model which better restores the actual galaxy parameters by only using unmasked dust-free regions of the galaxy.

The exact area to mask off is not easy to determine. A more extensive mask should better cover the regions of
a galaxy affected by dust and should lead to less biased fits. On the other hand, a larger mask means less of the
galaxy is actually used for fitting. In addition to that, the outer regions of galaxies are faint and have lower signal-to-noise ratios when compared to the inner galaxy regions. Thus, redundant masking is expected to make a fit less reliable. Moreover, the outermost regions of a galaxy can be dominated
by other structural components, such as thick discs and halos. Therefore, masking the central galaxy region can
switch the fitting, for example, from a thin disc to a thick disc and completely change the fitting results. To ascertain if there
is some trade-off between our intention to better cover dust attenuated regions and, at the same time, to use as much of the galaxy data as possible, we use several different masking strategies and compare their effects on the ability to recover the true parameters of galaxies.

\begin{figure}
  \label{fig:masking_examples} \centering
  \includegraphics[width=\columnwidth]{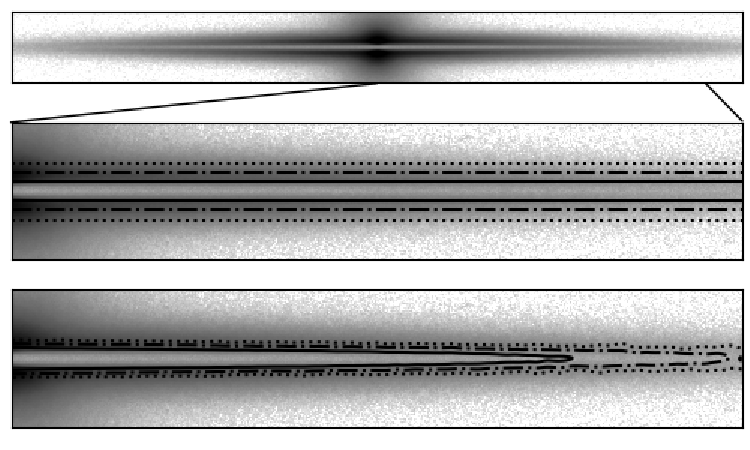}
  \caption{Top panel: an $r$-band image of a galaxy model with a face-on optical depth $\tau=1$. Middle panel: the same model
    enlarged with lines showing areas that would be covered by a flat mask with various $h_{\mathrm{mask}}$ values (see text): 2, 4,
    and 6 for solid, dot-dashed, and dotted lines respectively. Bottom panel: the same but with lines showing areas that would be
    covered by a relative mask for various $f_{\mathrm{mask}}$ values: 0.1, 0.25, 0.5 for dotted, dot-dashed and solid
    lines.}
\end{figure}

The simplest approach is to mask a narrow strip of a fixed height along the galaxy mid-plane. This mask has only one
parameter, its height $h_{\mathrm{mask}}$ that can be varied to govern the fraction of the galaxy area to be masked. In
order to link the height of the mask to the internal model parameters, we decided to measure the height in units of the dust component vertical
scale $z_{\mathrm{dust}}$. Hereafter, we refer to this simple mask as the ``flat'' mask. An application of such a flat mask
to a model image is demonstrated in the middke panel of Fig.~\ref{fig:masking_examples} for $h_{\mathrm{mask}}$ values of 2, 4, and 6.

\begin{figure}
  \label{fig:dust_vertical_slice} \centering
  \includegraphics[width=\columnwidth]{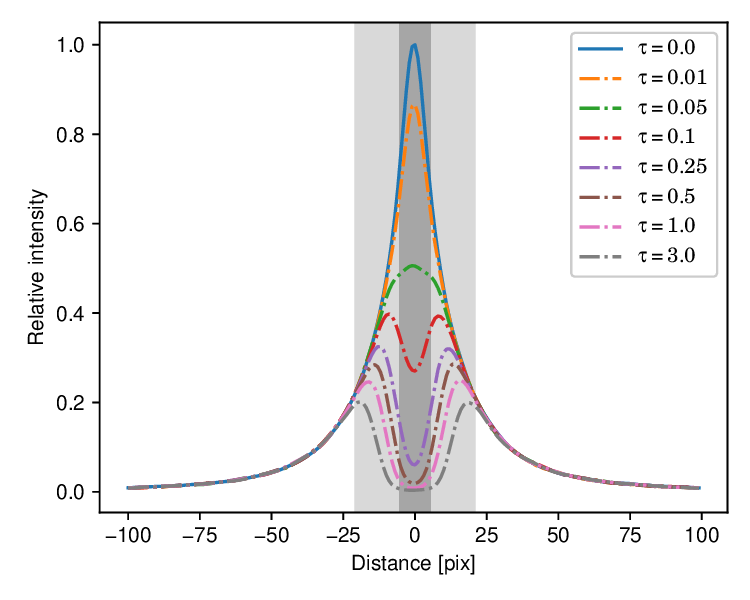}
  \caption{The impact of dust on the light distribution in a galaxy mock image illustrated by a vertical slice made
    through the image centre. Solid line: a dust-free ($\tau = 0$) model. The dot-dashed lines show the set of models with increasing
    values of $\tau$. The darker shaded region shows a distance of $z_{\mathrm{dust}}$ from the model mid-plane, a
    lighter one -- a distance of $4z_{\mathrm{dust}}$.}
\end{figure}

The appropriate size of the flat mask depends on the absorption strength. Even with the fixed values of
$z_{\mathrm{disc}}$ and $z_{\mathrm{dust}}$, the area where the dust impact is significant depends on the value of
$\tau$. This is illustrated in Fig.~\ref{fig:dust_vertical_slice} by a vertical slice made through the center of an image
for a dust-free model and for a set of models with an increasing value of $\tau$. It can be seen from the figure that,
while for galaxies with a relatively small dust content a flat mask with $h_{\mathrm{mask}}=1$ (i.e. height of a mask equal to $z_{\mathrm{dust}}$) may cover the affected areas of the image well enough, galaxies with prominent dust lanes require that the mask should be several times
wider. We also note that from this figure it becomes clear that even relatively transparent discs in a face-on orientation
(with $\tau=0.1$) demonstrate a prominent dust lane in an edge-on orientation; a central peak on the slice is
completely obscured by the dust and a darker depression is visible instead.

\begin{figure}
  \label{fig:mask_sizes} \centering
  \includegraphics[width=\columnwidth]{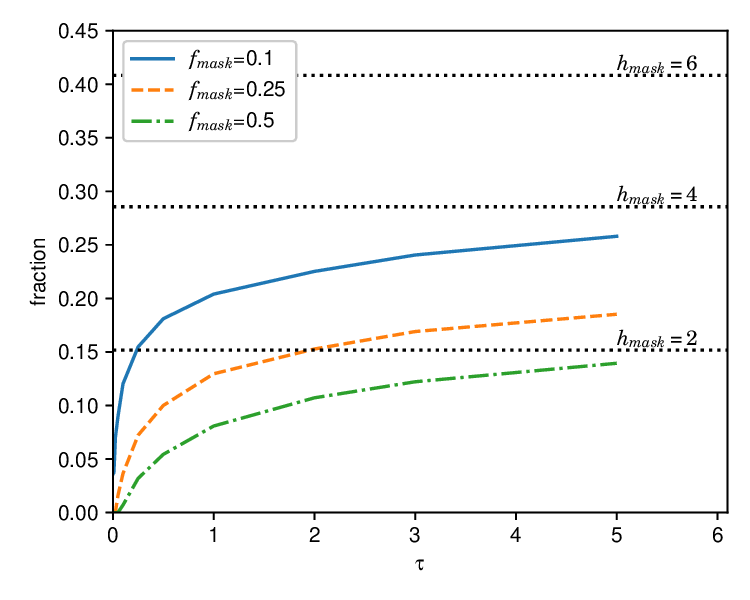}
  \caption{Fraction of the galaxy covered by masks of different types as a function of the dust disc optical depth $\tau$.
  Horizontal dotted lines show three flat masks, and the three curved lines show the behaviour of relative masks.}
\end{figure}

The drawback of using the flat mask is that it covers the mid-plane of a galaxy evenly for all radial distances from the galaxy centre, whereas most of the attenuation happens in the central region of a galaxy and decreases towards the periphery. This means that a mask that has a larger
height in the central region of a galaxy and becomes thinner toward the galaxy edges would more efficiently cover the
dust-affected regions of the galaxy. The exact shape of such a mask is not easy to find, as it
depends on the complex interplay between the parameters of the stellar and dust components. Luckily, when we work with
simulations, it is possible to determine the optimal parameters of such masks numerically. By comparing a mock image of a modeled dusty galaxy to an image of a model with the
same stellar components but without dust, we can find regions of the dusty model that are most affected by the dust
attenuation to mask them out. This leads to another masking strategy (we will call it a ``relative'' mask): a mask that
covers the regions where the relative change between the models with and without dust is higher than a given threshold. The
relative mask likewise has one free parameter to vary, the relative change between the two models $f_{\mathrm{mask}}$ above which
 we start our masking (in other words, a relative mask with $f_{\mathrm{mask}}=0.5$ covers regions where the dust
attenuation is higher than 50\%). The areas that are covered by a relative mask with $f_{\mathrm{mask}}$ values of 0.1,
0.25, and 0.5 are shown in the bottom panel of Fig.~\ref{fig:masking_examples}. It can be seen that the
relative mask covers the dust-affected regions of a galaxy more effectively: it is wider near the galaxy centre and
becomes thinner outwards. Another illustration of a comparison between these two approaches for creating dust masks is shown in
Fig.~\ref{fig:mask_sizes}. This figure shows the fraction of the total galaxy area (defined here as an image region that
contains 99\% of the total model flux) covered by different masks as a function of $\tau$. Since flat masks do not depend
on $\tau$, instead only depending upon $z_{\mathrm{dust}}$, their covered area appears as a flat strip. Relative masks, in contrast, depend on $\tau$ and grow as
the absorption increases. From this figure it is clear that the relative masking method is a more efficient way to cover the dust lane in terms of
the fraction of the galaxy image that is left for the upcoming fitting. As it will be shown later in Sec.~\ref{sec:simulations}, this leads to considerably better fitting results in terms of the precision of the recovered galaxy parameters.

Although the relative masking approach appears to be more promising, such a mask can be created easily only in the controlled conditions of a numerical experiment. In practice, one cannot readily find the relative fraction of light absorbed by
dust for every pixel of a galaxy image. To make this approach applicable to conditions where there is no such
information available (i.e. in real observations), we decided to train a neural network to produce a relative dust mask based on optical images of a galaxy.

\subsubsection{Neural networks for mask creation}
A relative mask represents a binary image where pixels with a value of 1 define a masked region in the
corresponding galaxy image and pixels with a value of 0 define the unmasked region. A generation of such masks from galaxy
images in several optical bands is the semantic segmentation problem. To tackle it, we employ a U-Net
\citep{2015arXiv150504597R} based network that was successfully used in \citet{2023MNRAS.519.4735S} to solve the Galactic cirrus segmentation problem
in SDSS optical bands.

The family of U-Net-based neural network models has been applied in various fields of science. These network
architectures are employed in medicine \citep{2017arXiv171205053I, Ching142760, 2018SPIE10581E..1BI,
  Andersson2019SeparationOW, Nazem20213DUA}, biology \citep{Kandel2020PhaseIW}, satellite image analyses
\citep{2017arXiv170606169I}, and astronomy \citep{2019MNRAS.484.5771A, 2021A&A...647A.120B, 2021MNRAS.505.3982B, 2021ApJS..254...41W, 2021MNRAS.503.3204V, 2022A&A...659A.199R, 2023A&A...669A.120Z}.

As the name suggests, the U-Net network model consists of two opposite paths. The down-sampling part, often called the
encoder, is used to capture features from an input image. The encoder consists of several repeated blocks of convolution and
max-pooling operations like in a typical convolutional neural network. The up-sampling part, often called the decoder, is used to
get precise localisations. The decoder also consists of repeated blocks of an up-sampling (increasing the resolution) of the feature map followed by convolution operations. Therefore, the spatial resolution of the tensor processed in the decoder increases. To get a localisation,
the features from the encoder are concatenated with the up-sampled features from the decoder via skip connections.

% We used a model that was successfully used in the problem of segmentation of galactic cirruses by optical images In
% general, the U-Net architecture consists of two symmetrical paths: an encoder to capture context and a decoder to get
% precise localisation. The encoder follows the typical architecture of a convolutional network with repeating
% convolution and max-pooling operations. Every step in the decoder consists of an upsampling of the feature map
% followed by a convolution. Thus, the decoder increases the resolution of the output. To get localisation, the features
% from the encoder are combined with the upsampled features from the decoder via skip connections.
%
% Originally, U-Net was proposed for biomedical image segmentation. This type of network architecture is successfully
% applied to various scientific and applied tasks such as medical image analysis \citep{2017arXiv171205053I,
% Ching142760, 10.1117/12.2293000, 2018SPIE10581E..1BI, Andersson2019SeparationOW, Nazem20213DUA}, cell biology
% \citep{Kandel2020PhaseIW}, and satellite image analysis \citep{2017arXiv170606169I}. It is also used in astronomical
% applications such as denoising, enhancing astronomical images \citep{2021MNRAS.503.3204V}, and stellar spectrum
% normalization \citep{2022A&A...659A.199R}.

\begin{figure}
	\label{fig:nn_architecture}
    \begin{center}
        \includegraphics[width = 0.48\textwidth]{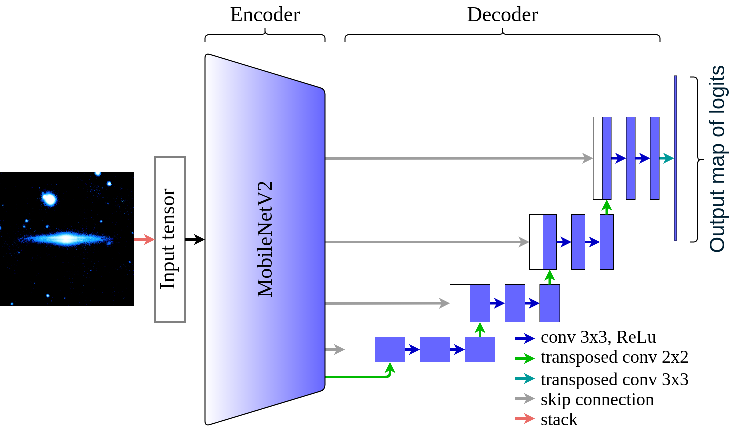}
    \end{center}
    \caption{The encoder-decoder architecture used in this study.}
\end{figure}

Our neural network model is implemented in the \texttt{TensorFlow2.x} framework \citep{tensorflow2015-whitepaper}. The
key difference between our solution and the original U-Net architecture is the encoder. As the encoder, we used the MobileNetV2
network model \citep{2018arXiv180104381S}, which is more lightweight than the original U-Net encoder, but has demonstrated a similar
performance in the Galactic cirrus segmentation problem \citep{2023MNRAS.519.4735S}. Fig.~\ref{fig:nn_architecture} displays the encoder-decoder architecture used.

\begin{figure*}
  \label{fig:nn_result}
  \includegraphics[width=0.8\textwidth]{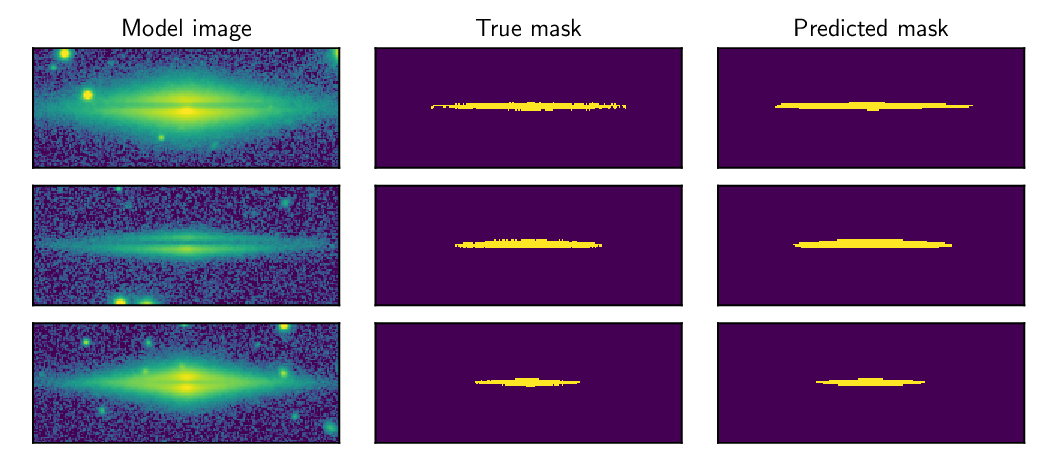}
  \caption{An example of using the neural network model for three simulated edge-on galaxies. Left panels: images of
    simulated galaxies in the $r$-band; middle panels: the corresponding relative mask ($f_{\mathrm{mask}} = 0.3$) computed from
    the model; right panels: masks generated by the neural network model. Background stars on the model images are from
    random SDSS fields (see text).}
\end{figure*}

During training experiments, we found accurate neural networks for different relative masks (ones with various $f_{\mathrm{mask}}$ values). We created separately trained networks for a set of $f_{\mathrm{mask}}$ values equal to 0.1, 0.3, 0.4, and 0.5. The data preparation to train the neural networks and the results of training experiments are described in Appendix~\ref{sec:appendix_neural}.

To find the best neural network to reproduce each relative mask, we use the IoU metric for the
masked regions:
\begin{equation} \label{eq:IoU}
	\mbox{IoU~} = \frac{\mbox{TP}}{\mbox{TP} + \mbox{FP} + \mbox{FN}}\,,
\end{equation}
where TP is the number of true positive pixel results where the network correctly predicts the masked pixel, FP is the
number of false positive pixel results where the network predicts the masked pixel but it belongs to the unmasked region, FN
is the number of false negative pixel results, where the network predicts the unmasked pixel but it belongs to the masked
region.

As demonstrated in Fig.~\ref{fig:nn_result}, the relative mask generated by our network quite accurately reproduces the
original relative mask. Note that for galaxies which are aren't viewed perfectly edge-on and where the dust lane is
shifted with respect to the galaxy center due to the projection effects, the mask generated by the neural network is also shifted and bent accordingly (see
the second row in Fig.~\ref{fig:nn_result}).

Quantitative results for different networks and training methods are shown in Table~\ref{tab:nn_metrics}. We summarise the results of our experiments as follows.
\begin{enumerate}
	\item As one can see in Table~\ref{tab:nn_metrics}, the best networks for all considered relative masks have a similar performance ($0.838 \leq$ IoU $\leq 0.847$).
	\item Networks trained using the \guillemotleft fine-tuning\guillemotright\,strategy demonstrate the best IoU per generation of all relative masks excluding a relative mask with $f_{mask} = 0.4$, but, as one can see in Table~\ref{tab:nn_metrics}, the advantage of these networks over ones trained from scratch or using the \guillemotleft transfer
learning\guillemotright\,strategy is insignificant.
	\item Our networks generate relative masks for a thousand galaxies in about 80 seconds when running predictions on an AMD Ryzen 9 3900X 12-Core CPU and about 40 seconds when running on an NVIDIA GeForce RTX 3060 GPU. 
\end{enumerate}

\subsubsection{Model with a dust component}
\label{sec:dust_model}
The second approach in accounting for the dust's impact during the decomposition process that we test in this work is
modifying the fitting model such that it includes a dust component. The correct treatment of this problem
requires heavy and time consuming computations both of light absorption and of scattering by dust grains. Moreover, such
computations are often based on a Monte-Carlo approach, so they introduce some randomness into the model
computations. This can impede using minimisation techniques based on gradient computations which are often used in
decompositions.

One simplification that can be made is neglecting light scattering, so that the only cause of losing photons is absorption. While scattered light can be important in disc galaxies, especially for near face-on orientations
\citep{Byun1994, Baes2001, Gadotti2010}, simulations show that for near edge-on orientations, the fraction of scattered
photons in the observed flux declines \citep{Pierini2004}. If a photon is scattered vertically in the disc, there is a high
probability it leaves the galaxy and cannot be observed from an edge-on orientation. If a photon is scattered along
the disc plane, where the optical depth is high, it will most likely experience another interaction with the dust to being either
absorbed or scattered away from the disc plane.

Under these conditions, the model flux at a given pixel $(x, y)$ can be found as a line-of-sight integral which includes
the optical depth term: for each point along the line-of-sight, we need to integrate over the dust density between this point
and the observer's position to compute the absorption for photons emitted at this point:
\begin{equation}
  \label{eq:los_integral_absorbed}
  I(x, y) = \int_{-\infty}^{+\infty} \rho_{\mathrm{stellar}}(s) e^{-\int^s_{-\infty}\kappa\rho_{\mathrm{dust}}(s')ds'}\, ds,
\end{equation}
where $\rho_{\mathrm{stellar}}(s)$ is the total luminosity density of a stellar model at a given point along a line-of-sight,
$\rho_{\mathrm{dust}}(s)$ represents the dust density with the extinction coefficient $\kappa$, with the negative line-of-sight direction being towards the observer. In this
case, the dust term accounts for the decrease in observed photons due to both absorption and scattering away from the plane of the disc.

To implement this approach, we modified the {\small IMFIT} code, by adding a new component function
that represents a combined model with a disc, bulge, and dust. The necessity to compute a double integral for every pixel of an image is a drawback for this method, since it imposes a high computational cost on the decomposition. On the other hand, it is more physically realistic than, for instance, a disc with a negative flux that was used to model a dust lane in \citet{Savchenko2017} and \citet{Smirnov2020}. Another advantage of implementing an approach with the direct integration in {\small IMFIT} is that there is no need in Monte-Carlo simulations in our computations. As a result, a Poisson noise that depends on the number of photon packages is not introduced in the results. Therefore, the output model image is smooth and can be compared with the input galaxy image using standard minimization techniques that involve computations of the numerical derivatives of $\chi^2$ (such as the Levenberg–Marquardt algorithm). Images obtained via Monte-Carlo simulations are noisy and different realizations of the same model can have slightly different $\chi^2$ values which impedes a gradient computation in the fitting procedure, and some other minimization technique is required (such as a genetic algorithm which does not rely on gradient computations but takes a lot more computational time). Using an AMD Ryzen 7 3700X 8-Core Processor, it takes less than 10 seconds to obtain a model image of a dusty galaxy with a size of $500\times 500$ pixels using our modified {\small IMFIT} code.

\section{Results of simulations}
\label{sec:simulations}

In this section we present the results of our simulations. To demonstrate how the dust distorts the measured values of the decomposition parameters, we make plots where the measured value of a parameter is plotted against the face-on optical depth $\tau$. These plots contain both decomposition results without dust correction and those obtained with using different
strategies to account for the dust (see Sec.~\ref{sec:decomp_dust_corr}), so that their outcomes can be compared. Similar simulations made by \citet{Gadotti2010, Pastrav2013a, Pastrav2013b} were performed for $\tau$ values up to 8, but the recent study by \citep{Mosenkov2018} shows that the total face-on absorption in disc galaxies does not reach such high values; the mean measured value for their sample of seven edge-on galaxies was found to be around $\tau=1$ in the $V$ band and the highest measured value was 2.01 for NGC~5907. On the other hand, galaxies with a higher amount of dust may exist, so we test our approaches of taking the dust into account in most extreme conditions with different possible $\tau$ values. Therefore, we decided to increase the investigated range of $\tau$ values well above the observed $\tau=2.01$ and set $\tau=5$ as an upper limit for our computations.

Also, we explore how different galaxy models (for example, with different bulge-to-total luminosity ratios or different relative stellar-to-dust disc scale
heights) are affected by the dust component with various optical depths.

Before proceeding to the results of simulations, we need to mention that the obtained discrepancies between the true values of the parameters and the ones that we infer via decomposition actually have two origins. The first is obviously the influence of dust, whereas the second is some intrinsic decomposition biases. As was previously found in \citet{Gadotti2010, Pastrav2013a, Pastrav2013b}, even if dust is absent in a model, the measured decomposition parameters can differ from their true input values. In those studies, this difference was attributed to a mismatch between the models. While the radiative transfer model used for the image creation was three-dimensional, the decomposition model contained a simple two-dimensional exponential disc. This two-dimensional model cannot take the disc's vertical structure into account, which results in an increase of the disc's inclination. For the edge-on galaxies where the disc thickness plays a dominant role, the two-dimensional exponential model cannot be applied to infer disc properties, since the projected light distribution in this case is not exponential, but can be described as a combination of a Bessel  and hyperbolic secant function \citep{vanderKruit1981}.

In this paper, we also find that the results of the decomposition via dust-free models do not exactly match the values of the input parameters, especially for the bulge component. A possible explanation for this error is the presence of the noise we added intentionally to our mock images. This noise has two sources. The {\small SKIRT} code operates in terms of photon packages that are emitted inside the galaxy and then propagate through the galaxy body towards the observer. Since the number of such packages is finite, a model image is not smooth, but demonstrates some photon noise. The second source of the noise is the one we added to convert a SKIRT simulated image to an ``observed'' image. In this work, we model images of the $r$-band of the SDSS survey, therefore, we use the noise characteristics of this instrument (see Sec~\ref{sec:synthimag}).

To examine how these two noises impact our decomposition results, we run a number of fits with different noise parameters. First, to study how the number of photon packages in a SKIRT simulation affects the decomposition quality, we create the same model but with $10^6$,\, $5\cdot 10^6$, and $10^7$ photon packages and then decompose these models using the same technique. To inspect the impact of the included camera noise, we run these three simulations with a different number of photon packages twice: with and without adding the camera noise. The results of these experiments are listed in Table~\ref{tab:noised_decompositions}, where the measured value of the S\'ersic index is shown along with the true value of 4.0. One can see that the measured value is almost unaffected by the number of photon packages, while adding the camera noise makes a significant change to the retrieved parameters. The fact that even for a noise-free model we do not recover the correct value of the S\'ersic index, but have a somewhat lower value, probably originates from an interference between the bulge and disc components that overlap in the image which leads to a degeneracy of their parameters.

From these simulations, we make a conclusion that the added camera noise is the main source of error when decomposing a dust-free galaxy image, and that all follow-up experiments with dusty models also include this bias. Since our main goal is to simulate the decomposition errors for real observations (which always contain noise), we do not correct our results for this bias, but emphasise that the estimated decomposition errors can have various reasons apart from the dust impact which we discuss in detail in the next sections. All simulations in the subsequent sections are made for the SDSS $r$ band.

\begin{table}
  \centering
  \begin{tabular}[h]{lccc}
    \hline
    Photon packages & $10^6$ & $5\cdot 10^6$ & $10^7$ \\
    \hline
    With camera noise  & 3.46  & 3.44  & 3.44 \\
    Without camera noise  & 3.88  & 3.86 & 3.86 \\
    \hline
  \end{tabular}
  \caption{Measured bulge S\'ersic index for a set of decompositions with different noise parameters.}
  \label{tab:noised_decompositions}
\end{table}

\subsection{Dust impact on the bulge parameters}
\label{sec:dust_impact_bulge}
The bulges of disc galaxies show a peak intensity at their centres and, generally, a rather swift decrease in their surface brightness in
their outer regions. As shown in \citet{Gadotti2010}, even for galaxies with inclinations far from an edge-on
orientation, the bulge parameters can be strongly affected by dust. To demonstrate how the dust component affects
the fit parameters of the bulge in an edge-on orientation, we run our algorithm for three models with different bulge parameters: a big, medium, and small bulge (their parameters are listed in
Table~\ref{tab:three_bulges}). We note that all models have the same value of the S\'{e}rsic index equal to 4 (i.e. they represent de Vaucouleurs bulges). Bulges with different S\'{e}rsic indices have different concentrations and, thus, it is natural to expect that the presence of dust affects their measurements differently, but in this article we do not consider this problem. The disc parameters of these models are the same as in the basic model
(Tab.~\ref{tab:basic_model}).

\begin{table}
  \centering
  \begin{tabular}[h]{crrrr}
    \hline
    Model  & $r_e$ [pc] & $n$ & $q$ & $B/T$ \\
    \hline
    Small  & 700  & 4  & 0 & 0.1 \\
    Medium & 900  & 4  & 0 & 0.2 \\
    Big    & 1500 & 4  & 0 & 0.3 \\
    \hline
  \end{tabular}
  \caption{Parameters of the three models corresponding to the different bulges.}
  \label{tab:three_bulges}
\end{table}

\begin{figure*}
  \label{fig:bulge_no_corrections} \centering
  \includegraphics[width=0.8\textwidth]{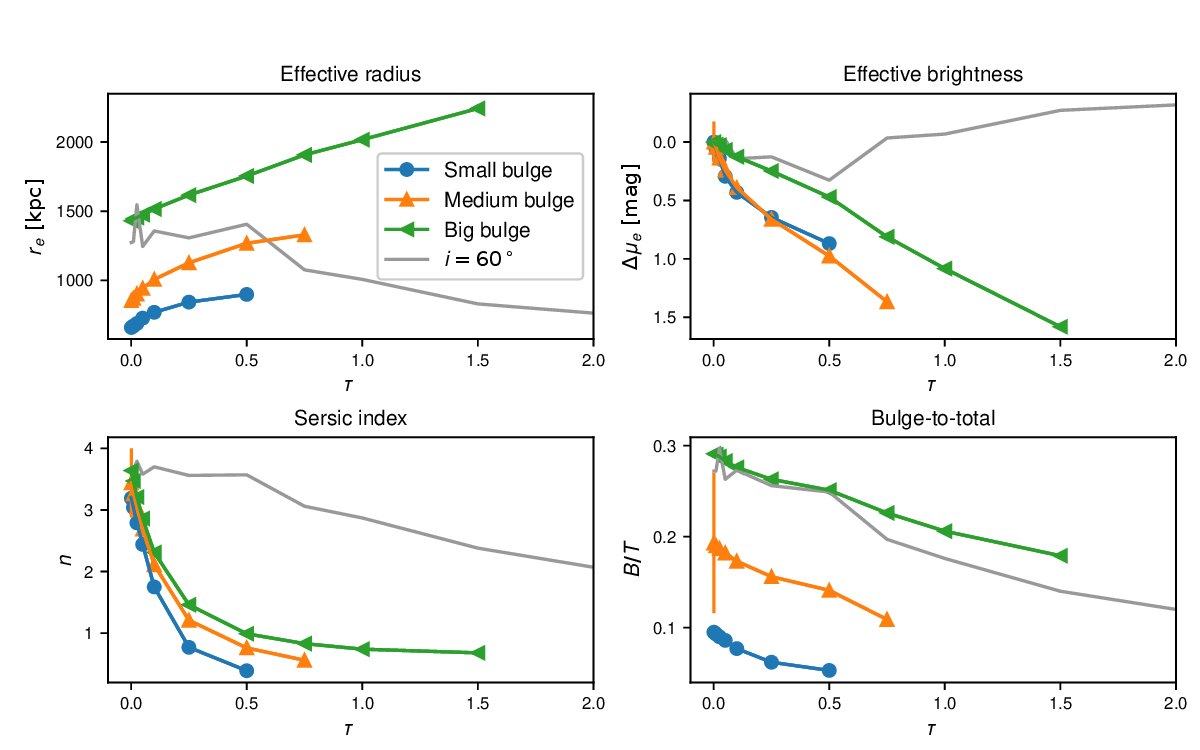}
  \caption{The dependence of the measured bulge parameters' values on the face-on optical depth $\tau$ for the effective
    radius, effective surface brightness, S\'{e}rsic index, and bulge-to-total luminosity ratio. Three bulge models are shown: a
    ``small bulge'' model ($r_e=700$~pc, $B/T=0.1$), ``medium bulge'' model ($r_e=900$~pc, $B/T=0.2$), and ``big bulge''
    model ($r_e=1500$~pc, $B/T=0.3$). The grey line shows the results for a ``big bulge'' model with an inclination
    of 60 degrees.}
\end{figure*}

Fig.~\ref{fig:bulge_no_corrections} demonstrates how the measured values of the bulge's effective radius, effective surface
brightness, S\'{e}rsic index, and bulge-to-total luminosity ratio depend on the amount of dust in the model. From this figure, we note that when a galaxy is viewed edge-on, the bulge parameters deteriorate very quickly as the optical depth of the dust
component increases. Although we run our simulations for a range of $\tau$ from 0 to 5.0, in all three models, the bulge component
begins to diverge long before reaching the maximal value of $\tau$. For example, the \textit{small bulge} model
collapses to the lower limit of the S\'{e}rsic index at $\tau \approx 0.5$; after that point the gradient descent algorithm
starts to converge to random values around the initial conditions. This indicates that the bulge is obscured to such a great degree that it
does not affect the total value of the $\chi^2$ statistics. The same happens for the \textit{medium bulge} model at
$\tau \approx 0.75$ and for the \textit{big bulge} model at $\tau \approx 1.5$. Since all the bulge models have collapsed by $\tau \approx 1.5$, we do not show the results of modelling higher absorptions in this figure.

For smaller values of $\tau$, where we managed to obtain at least some measurable values for the bulge parameters, their
behaviour is similar in each model. The fit effective radius of the bulge grows with optical depth, which can be easily understood from a geometrical point of view: the central peak of the bulge is obscured, but its outer regions (outside of the plane of the disc) are essentially unaffected. Thus the radius where half of the total observed bulge flux is confined must be larger than in the
dust-free case. The observed S\'{e}rsic index decreases for the same reason: the obscured central peak leads at a
flatter apparent surface brightness distribution, i.e. lower values of $n$. For the \textit{small bulge} model, the measured
value of the S\'{e}rsic index drops from a true value of 4.0 to a value of 2.0 already at $\tau \approx 0.075$, which value of $\tau$ corresponds to when the
dust component begins to appear as a darker lane in the galaxy image (see Fig.~\ref{fig:dust_vertical_slice}). Therefore,
even if a visual inspection does not reveal a dust lane in a galaxy, it can still contain enough dust to render the parameters
of a small bulge completely distorted. This in turn affects some standard galaxy scaling relations, which contain the parameters of the bulges (e.g. the Kormendy relation), or, for example, make the ``classical bulge -- pseudobulge'' dichotomy less pronounced. 

The other two panels in Fig.~\ref{fig:bulge_no_corrections} show the results for the effective surface brightness (in
terms of its difference from its true value) and the observed bulge-to-total luminosity ratio. There is no surprise that for
models with a higher dust content, our measurements show progressively fainter bulges, although for the bulge-to-total luminosity ratio,
 the changes are not as extreme as for the S\'{e}rsic parameter. This happens probably because the disc component is also obscured
by the dust and this suppresses the total $B/T$ ratio shift to some level.

For comparative purposes, Fig.~\ref{fig:bulge_no_corrections} also displays the results of our simulation for the \textit{big bulge} model inclined by 60 degrees. The result of this simulation closely follows the results reported in \citet{Gadotti2010} where this inclination was the highest among those they considered: the effective radius, S\'{e}rsic index, and bulge fraction all decrease with optical depth (see their figures 3 and 4), which validates our simulations. It is clear from this comparison that for the edge-on orientation, the impact of the dust on the bulge parameters is disparately higher than for mildly inclined galaxies. We note that the difference between the edge-on and non-edge-on cases is not solely quantitative, but a contrasting behaviour can be observed instead. For the edge-on models, the measured value of the bulge effective radius tends to be higher then the true one, whereas for a mildly inclined model it becomes lower. The same behaviour was also confirmed for models with exponential and de Vaucouleurs bulges \citep{Pastrav2013b} for a wide range of inclinations, from face-on to almost edge-on, but for decomposition with infinitely thin discs.

\begin{figure*}
  \label{fig:bulge_corrected} \centering
  \includegraphics[width=0.9\textwidth]{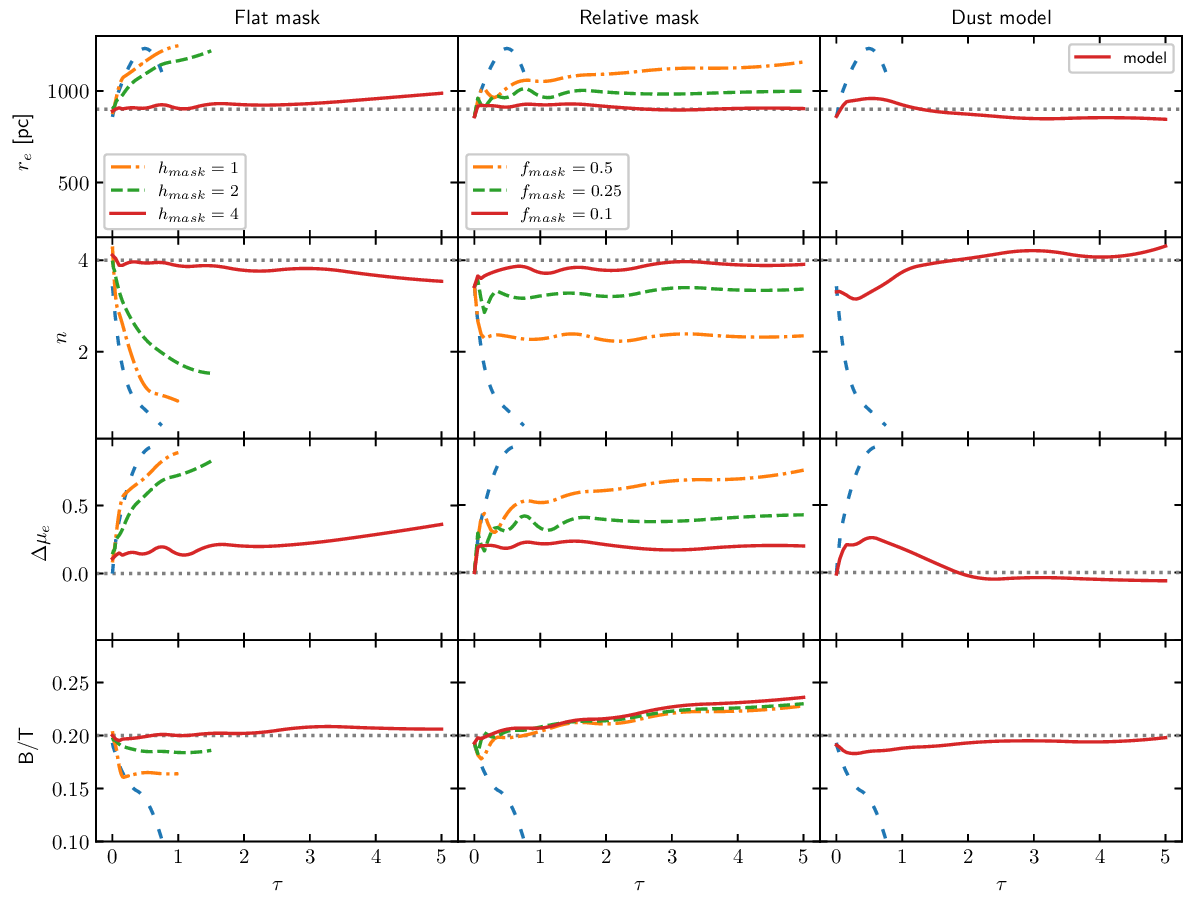}
  \caption{Results of fitting with dust correction methods for the bulge parameters. Each panel
    shows the measured value as function of $\tau$. The top row shows results for the effective radius $r_e$, the second
    row shows the S\'{e}rsic index $n$, the third one shows the error of the effective surface brightness $\Delta \mu_e$, and the
    bottom row shows the measured bulge-to-total luminosity ratio.  Different dust correction techniques are shown in the columns: the
    leftmost column shows the decomposition with flat masks with $h_{\mathrm{mask}}=1$ (dot-dashed line), $h_{\mathrm{mask}}=2$
    (green densely dashed line), and $h_{\mathrm{mask}}=4$ (solid line); the middle column shows relative dust masks
    with $f_{\mathrm{mask}}=0.5$ (dot-dashed line), $f_{\mathrm{mask}}=0.25$ (green densely dashed line), and
    $f_{\mathrm{mask}}=0.1$ (red line); the rightmost column shows the results of decomposition with a dust model (red
    line). For comparison, each panel shows the uncorrected values (blue sparsely dashed line) and the true values as a
    grey dotted level.}
\end{figure*}

Fig.~\ref{fig:bulge_corrected} shows the decomposition results for the \textit{middle bulge} model with the aid of various techniques for taking the dust into account. There are four bulge parameters shown in this figure (the
effective radius $r_e$, the S\'{e}rsic index $n$, the error of the effective surface brightness
$\Delta \mu_e=\mu_e^{\mathrm{measured}}-\mu_e^{\mathrm{true}}$, and the bulge-to-total ratio B/T) and three dust
correction methods (the flat mask, relative mask, and dust model), for a total of twelve panels. The columns show different
methods, the rows, different parameters. Each panel also contains results for the uncorrected decomposition (the same as
in Fig.~\ref{fig:bulge_no_corrections}) for comparison purposes, and the true value of each parameter marked as the horizontal dotted
line. Below we describe all three approaches for the dust correction separately. 

\subsubsection*{Flat mask}
The results of the decomposition with flat masks are shown in the leftmost column of Fig.~\ref{fig:bulge_corrected} for three
mask sizes: $h_{\mathrm{mask}}=1, 2, \mbox{and } 4$ (a yellow dash-dotted line, a green dashed line, and a red solid line, respectively), along with the results of the unmasked decomposition (the blue sparsely dashed line). As we discussed
earlier, without dust correction the bulge model collapses at $\tau \approx 0.75$. The same happenes for decomposition
with a relatively narrow flat mask: for a mask with $h_{\mathrm{mask}}=1$, the model collapses at $\tau \approx 1$, and for
$h_{\mathrm{mask}}=2$ at $\tau=1.5$. Therefore, the narrow flat mask provides almost no improvement compared to the unmasked case:
the decomposition results become highly distorted even for low values of $\tau$.

For a wider mask ($h_{\mathrm{}}=4$), the results are considerably better. Even though all three parameters still deviate
from their true values, this deviation is confined in a narrower range, and moreover, the model converges
successfully for all considered $\tau$ levels up to 5.0. There is still an apparent systematic shift in all three
parameters in the same direction as for the decomposition with narrower masks, but for a moderate dust content it is not
larger than typical uncertainties for decomposition results.

\subsubsection*{Relative mask}
The middle column in Fig.~\ref{fig:bulge_corrected} shows the results of the decomposition with three relative masks:
$f_{\mathrm{mask}}=0.5$, $f_{\mathrm{mask}}=0.25$, and $f_{\mathrm{mask}}=0.1$ (a yellow dash-dotted line, a green dashed line, and a red solid line, respectively). The first fact that should be noted is that the
results of the decomposition with the relative mask are better than both those without any masking and with a flat mask.

For a model with $\tau=1$, the total area that is covered by the relative mask with $f_{\mathtt{mask}}=0.5$ is almost the
same as the flat mask with $h_{\mathrm{mask}}=2$ (in fact, it is 10\% smaller, see Fig.~\ref{fig:mask_sizes}), so the
green curve in the left column can be directly compared to the yellow curve in the middle one to assess the
performance improvement of the relative mask. It is clear that despite having virtually the same total covered area, the
relative mask does a better job in reducing the dust impact on the decomposition results. The same is true for all tested
values of $\tau$: the relative mask allows us to recover the bulge parameters more reliably while covering a smaller fraction of the
galaxy image.

\subsubsection*{Dust model}
The results of applying our combined {\small IMFIT} model -- which contains a bulge, disc, and dust component --
to decompose a synthetic image are shown in the right column of Fig.~\ref{fig:bulge_corrected}. This
decomposition does not employ any masking, and taking the dust into account only happens by including the dust term during the
line-of-sight integration (Eq.~\ref{eq:los_integral_absorbed}) for the model. The figure suggests that the performance of
this approach is comparable with the results of the widest flat ($h_{\mathrm{mask}}=4$) and relative
($f_{\mathrm{mask}}=0.1$) masks.

Our inability to recover the precise values of the structural parameters when utilizing this approach can be attributed to two facts: i) it
does not include the light scattered by the dust, and ii) it reflects the general problems of a decomposition with a complex model, where a
degeneracy between some parameters can occur and lead to a systematical shift in parameters \citep{Gadotti2010}.

Although the dust model technique does not show considerably better results when compared with dust masking, whilst taking an
order of magnitude longer in computation time, it still has an advantage in that it does not lose information about the
galaxy by masking out some galaxy regions. If a galaxy has a more complex structure, such as two (thin
+ thick) stellar discs, the thin disc can be completely covered by a dust mask, whereas the dust model approach can still recover the disc structure with some accuracy. 

\subsection{Stellar discs}
\label{sec:dusty_discs}
Since the dust disc is embedded inside the stellar disc, it is natural to expect that the impact of dust on the fit disc parameters should depend on the scaling relations between the structural parameters of the two discs. To investigate this,
we ran simulations for a set of various dust disc parameters while keeping the stellar disc parameters fixed to those
given in the \textit{basic model} (Table~\ref{tab:basic_model}).

The first set of simulations regards a relation between the radial scale lengths for the stellar and dust
discs. Observations show that there is a correlation between the radial scale length of the emission profile at 3.4~$\mu$m (which traces the bulk of the stellar mass in a galaxy) and that at 100~$\mu$m (which is dominated by the emission of cold dust, \citealt{Mosenkov2022}, see also \citealt{2019A&A...622A.132M} for a similar correlation but for the effective radius). Moreover, \citet{Casasola2017} found for 18 face-on spiral galaxies the following ratio between the disc scale lengths of the stellar and dust surface density distributions $h_{\Sigma \mathrm{dust}}/h_{\Sigma \mathrm{star\,3.6}\mu \mathrm{m}}=1.80\pm0.13$ which is generally consistent with the results from radiative transfer modeling (Mosenkov et al. in prep.). In this study, we decided to consider three possible situations: a dust disc which is slightly shorter
than a stellar disc ($h_{\mathrm{dust}}=3000$~pc -- some galaxies harbor a shorter dust disc than their stellar disc, e.g. NGC\,4013, \citealt{Mosenkov2018}), both discs having the same radial scale ($h_{\mathrm{dust}}=4000$~pc), and a dust disc which is more extended
than a stellar disc ($h_{\mathrm{dust}}=5000$~pc), to see if these differences translate into various systematical shifts in
the derived parameters.

\begin{figure*}
  \label{fig:disc_no_corr} \centering
  \includegraphics[width=0.9\textwidth]{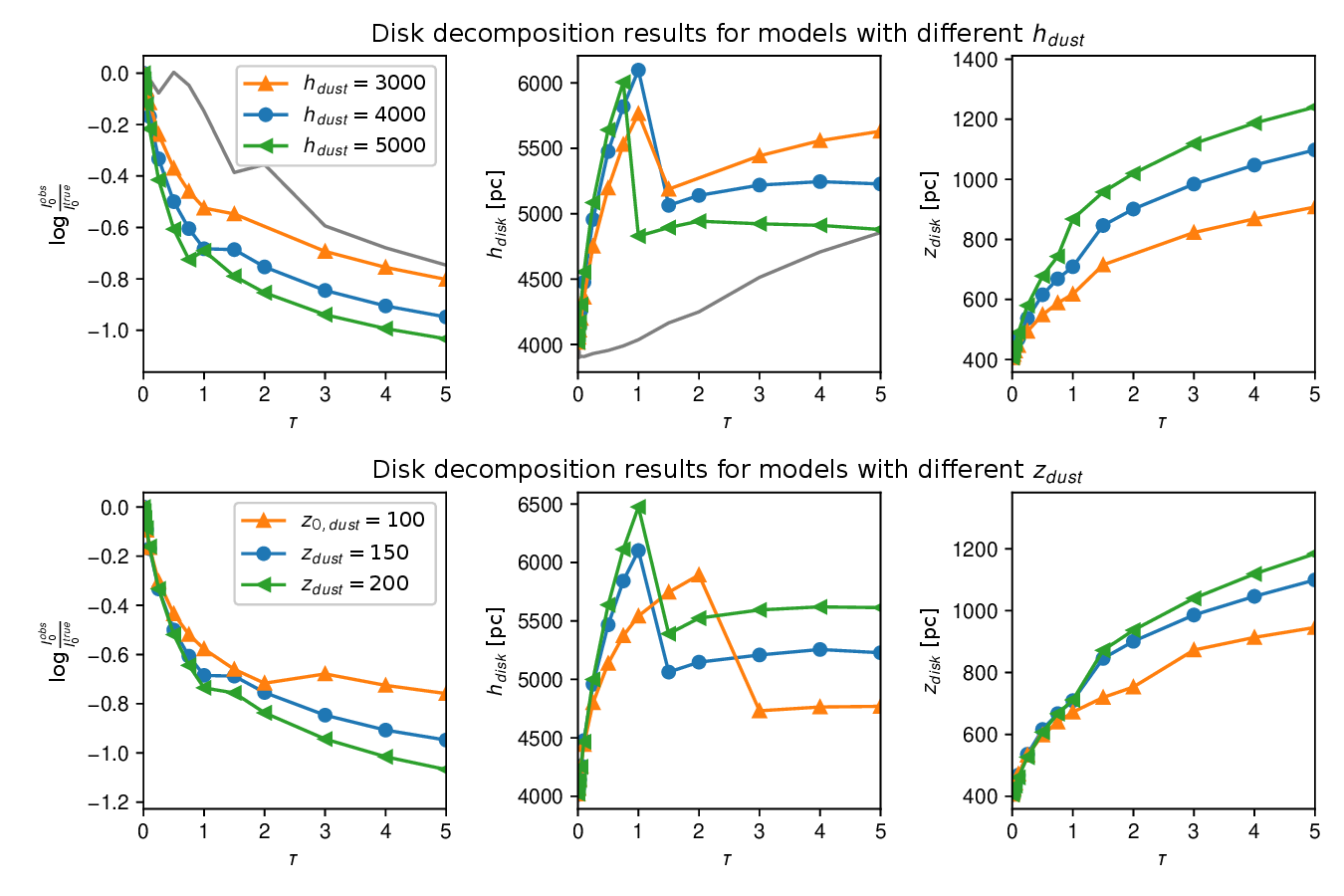}
  \caption{Disc parameters retrieved by decomposition for a set of models with different dust disc
    parameters. Top row: results for the disc central surface brightness (left panel), radial exponential scale
    (middle), and vertical scale (right) for three dust components with a radial scale shorter, equal to, and larger than that
    of the stellar disc. For comparison, the grey lines for the panels about surface brightness and exponential scale show
    results for the same disc with $h_{\mathrm{dust}}=4000$ pc, but with an inclination of 60 degrees. Bottom row: the same
    parameters obtained for dust models with different vertical scales ($z_{\mathrm{dust}}=$ 100, 150, and 200~pc).}
\end{figure*}

The results of these simulations are demonstrated in the top row of Fig.~\ref{fig:disc_no_corr}. The values of the measured
parameters are shown as a function of the face-on dust optical depth $\tau$. The left panel shows the observed central
surface brightness in terms of a decimal logarithm of a fraction of the observed to true value, without dust
attenuation.  The middle panel shows the measured radial exponential scale of the stellar disc. The right panel shows
the measured value of its vertical scale. For comparison, in the panels with surface brightness attenuation and radial scale we also show the results of simulations for a \textit{basic  model} inclined at 60 degrees. Again, the results of decomposition with the inclined disc, in general, follow the results presented in \citet{Gadotti2010}, and for the edge-on orientation, the  impact of the dust on the disc parameters is higher.

From these plots it is clear that the dust component severely changes the observed values of the stellar disc. All three models show a similar behaviour for the observed central surface brightness: its value drops quickly to about 20\%
the original level at $\tau \approx 0.5\div 0.75$, and then after a short pause a shallower attenuation is observed. A
possible explanation for this behaviour is that at these values for the face-on optical depth, the edge-on disc absorption exterminates almost all observed photons near the galaxy midplane, and then the following attenuation occurs in regions far
from the disc plane where the dust content is lower. From this plot it can also be seen that the general behaviour
of the attenuation in edge-on discs follows the what's seen in inclined discs, although in edge-on discs it is qualitatively stronger.

A different picture is seen for the disc radial scale length (top-middle panel). While for a disc
inclined at 60 degrees the fit value of this parameter gradually increases with $\tau$ and reaches a relative
change of 10\% only at $\tau \gtrsim 2$ (i.e. this disc would demonstrate a very prominent dust lane if observed in an
edge-on orientation), for an edge-on disc a very sharp increase is observed. In this case, at the same value of
$\tau \approx 1-1.5$, the relative change of the $h_{\mathrm{disc}}$ reaches a peak that is approximately 50\% higher than the true value (a similar $\approx 50\%$ increase of the observed values of $h_{\mathrm{disc}}$ was reported for an (almost) edge-on orientation, but for decomposition with an infinitely thin disc). After that, the observed
value of $h_{\mathrm{disc}}$ quickly drops to $\approx 5000$~pc (25\% higher than the true value), and its 
behaviour for higher $\tau$ depends on the relation between $h_{\mathrm{dust}}$ and $h_{\mathrm{disc}}$. If
$h_{\mathrm{dust}} < h_{\mathrm{disc}}$, the curve of the measured value for $h_{\mathrm{disc}}$ starts to grow again. If
$h_{\mathrm{dust}} \ge h_{\mathrm{disc}}$, it appears to be more stable and does not change appreciably for a wide range of $\tau$.

The observed break at $\tau \approx 1.5$ in the plot for $h_{\mathrm{disc}}$ (top-middle plot of Fig.~\ref{fig:disc_no_corr}) is caused by issues with the bulge fitting. For high absorption levels, the S\'ersic component cannot fit the bulge properly because the dark dust lane in the galaxy midplane suppresses its maximal brightness. This leads to the appearance of two under-fit bulge ``remnants'' above and below the disc plane, where the bulge protrudes from the dusty disc. Their fraction in the residuals increases with $\tau$ because the bulge fit becomes progressively worse and the disc itself becomes darker due to absorption, while these bulge regions remain almost unobscured. At some point, it becomes more efficient for the model (in terms of its achieved $\chi^2$ value) to fit these bulge remnants as part of the disc component. This leads to a more concentrated disc model and, therefore, a shorter radial scale.

The results for the last disc parameter, its vertical scale height $z_{\mathrm{disc}}$, are shown in the top-right panel of Fig.~\ref{fig:disc_no_corr}. Again, all three models show the same general trend in that the fit value of $z_{\mathrm{disc}}$ increases with  $\tau$. This is easy to understand since the dust absorbs more photons close to the
disc plane, making the vertical brightness distribution flatter, therefore it can be approximated with a higher value
of $z_{\mathrm{disc}}$. We also point out a clear systematic trend: the larger the dust exponential
scale height, the greater its impact on the observed $z_{\mathrm{disc}}$ value. Since in both \citet{Gadotti2010} and \citet{Pastrav2013a, Pastrav2013b}, an infinitely thin disc model was utilized during decomposition, the $z_{\mathrm{disc}}$ value was not inferred in their simulations and our results cannot be compared with these studies.

The bottom row of Fig.~\ref{fig:disc_no_corr} shows the results of our simulations with different $z_{\mathrm{dust}}$ values. One can see that, in general, the behaviour of all three parameters are the same as those 
obtained for models with varied $h_{\mathrm{dust}}$ values. We note that all else being equal, a thicker dust
disc leads to more distorted stellar disc parameters.

\subsubsection*{Flat mask}
\begin{figure*}
  \label{fig:disc_corrected} \centering
  \includegraphics[width=0.9\textwidth]{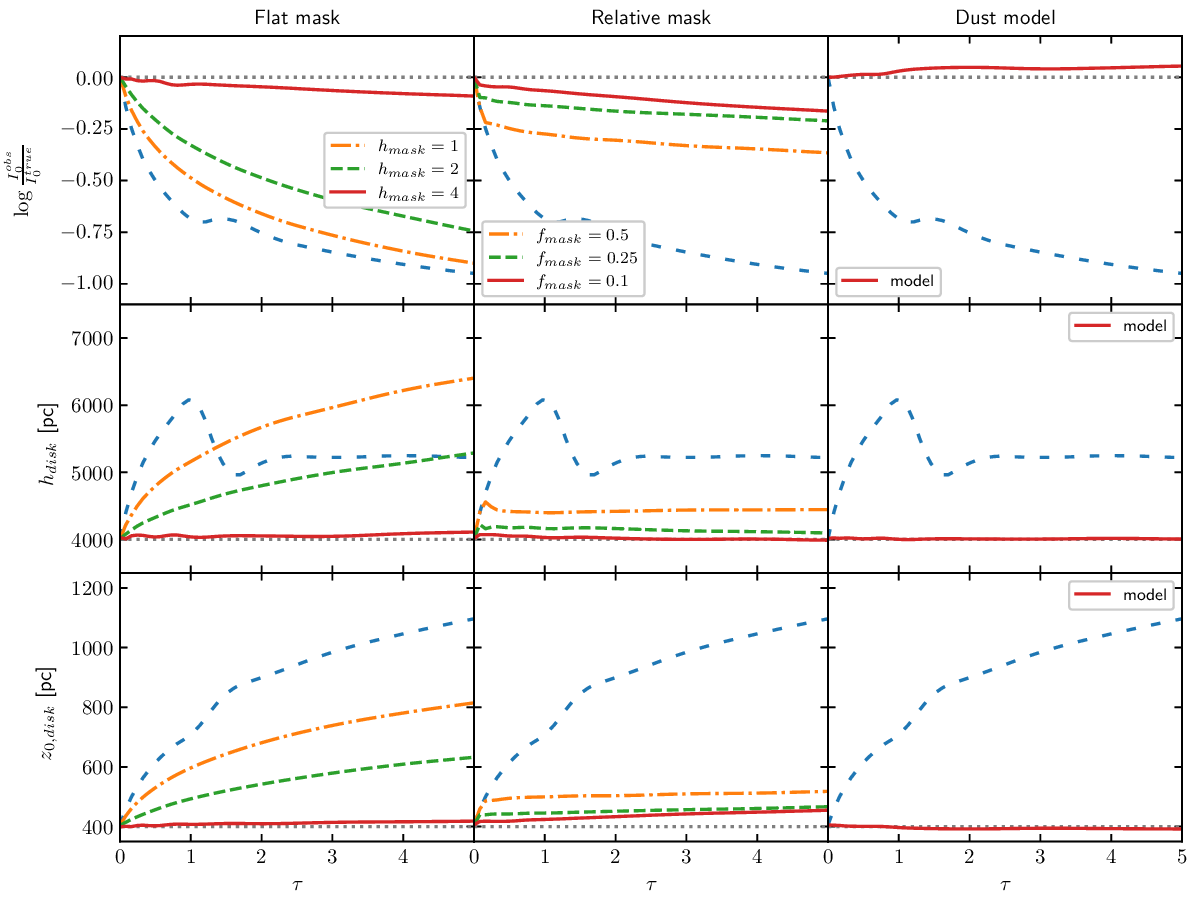}
  \caption{Results of the fitting with different dust correction methods for the disc parameters. The
    general structure of the figure follows similarly to the bulge decomposition results (see the caption of
    Fig.~\ref{fig:bulge_corrected}), except for the disc parameters, the rows correspond to the central surface brightness (top), the radial scale
    (middle), and the vertical scale (bottom).}
\end{figure*}

In this and the next two paragraphs, we describe the results of taking the dust into account using three various
techniques in a similar way as it was done for the bulge. We begin with a flat mask approach, which results are
demonstrated in the left column of Fig.~\ref{fig:disc_corrected}. The figure shows that while a flat mask allows one
to enhance the quality of the decomposition, only a mask that is four times wider than $z_{\mathrm{dust}}$ yields
parameter estimates that are close to their true values. Narrower masks result in a systematic shift in the parameters
(a rapid decline of the disc flux, and an increase in both the radial and vertical scales) that depend on the value of $\tau$.

\subsubsection*{Relative mask}
For retrieving disc parameters, a relative mask (middle column of Fig.~\ref{fig:disc_corrected}) provides better
results than a flat mask. Even the mask with the smallest covered area ($f_{\mathrm{mask}}=0.5$) gives estimates of the radial and
vertical disc scales closer to the true values when compared to the considerably larger flat masks. However, the central surface brightness is still systematically underestimated. The most extended relative
mask ($f_{\mathrm{mask}}=0.1$) allows us to recover the radial exponential scale almost perfectly, although for this value, there are some trends present in the central surface brightness and the vertical scale. Their errors are comparable to the
characteristic uncertainties of the decomposition. We conclude that, as is the case for the bulge parameters,
a disc decomposition with a relative mask generally gives better results for some fixed fraction of the covered galaxy image.

\subsubsection*{Dust model}
The right column of Fig.~\ref{fig:disc_corrected} shows the results of a decomposition using a model with a
dust component. This approach allows us to almost perfectly recover both the radial and vertical exponential scales, and there is
only a slight overestimation of the disc central surface brightness. The fact that our model allows us to retrieve almost exact values for the disc parameters even for a high dust content seems to confirm the fact that light scattering has little impact on galaxy decompositions in an edge-on orientation, as was mentioned by \citet{Pierini2004}. Also, for an the edge-on orientation, the spatial overlap between the disc and bulge components is lower, which reduces the degeneracy of their parameters.

\section{Demonstration on real galaxies}
\label{sec:real_galaxies}
In previous sections, we described the techniques we used to account for a dust component and the results of their
application to synthetic images of several model galaxies. To further validate these approaches in inferring
the structural parameters of disc galaxies, we present the results of decompositions for a couple of real
edge-on galaxies. We use two methods to account for the dust component: a neural-network generated relative dust mask, and an
{\small IMFIT}-model with a dust component. For this purpose, we selected two edge-on galaxies with prominent dust lines but without
significant complex features in their discs (warps, flarings, bright halos, etc.) from the EGIPS
catalog\footnote{\url{https://www.sao.ru/edgeon/catalogs.php?cat=EGIPS}}, which contains a sample of 16551 edge-on
galaxies. The selected objects are PGC\,27896 and PGC\,2441449. We downloaded their images from the Pan-STARRS1 survey
\citep{Chambers2016, Flewelling2020} in the $i$-band.

\begin{figure*}
  \label{fig:real_galaxies} \centering
  \includegraphics[width=0.8\textwidth]{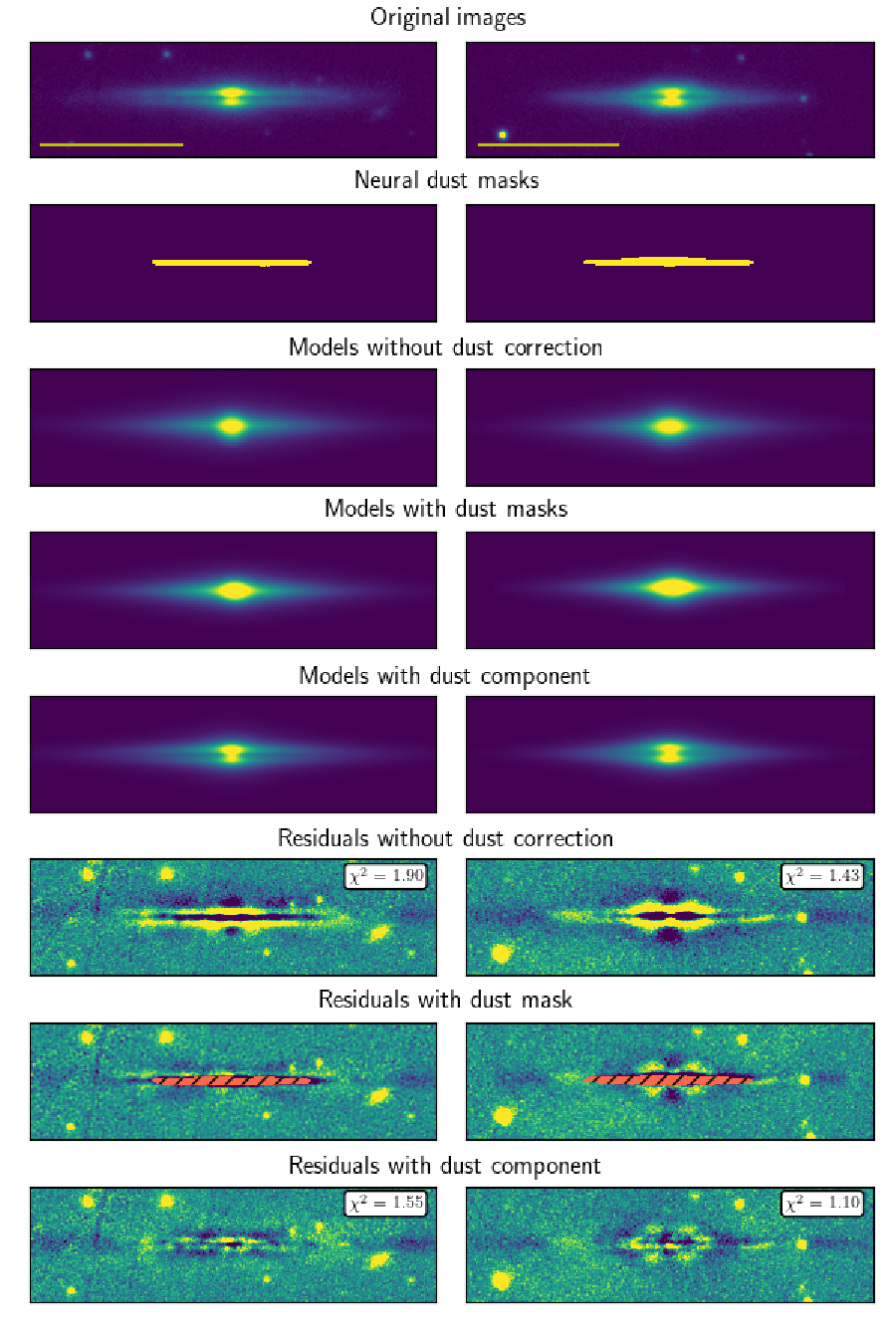}
  \caption{Decomposition of two real galaxies, PGC\,27896 (left column) and PGC\,2441449 (right column). Top panels: images
    of the galaxies in the $i$ band with 30\arcsec scale bars. Second row: dust masks created by our neural network. Third, fourth and
    fifth rows: best fit models for a decomposition without taking the presence of dust into account, with a dust mask, and
    with a dust model. Three bottom rows: residual maps for these decomposition approaches. Red hatched areas on the residuals with a dust mask show masked areas. The fit $\chi^2$ statistics is shown for decompositions without dust correction or with a dust component (the decompositions with the dust mask have a different masked area, so its statistics cannot be accurately compared with the other two methods).}
\end{figure*}

To compare decomposition results with and without dust correction, we first decompose these galaxies using a
photometric model consisting of a S\'ersic bulge and a 3D exponential disc, along with taking a dust mask or a dust model into account.
In all cases, we use the same PSF image that was obtained by fitting a Moffat function to a stacked image of bright
isolated stars in the galaxy frame. We masked off background and foreground objects using a segmentation map produced by the {\small
  SEXTRACTOR} package \citep{Bertin1996}. The results of our decompositions are shown in
Fig.~\ref{fig:real_galaxies}; the left column is for PGC\,27896, the right is for PGC\,2441449. The top panels show the
original images of these two galaxies in the $i$ band with yellow bars marking a 30\arcsec scale.

The second row in Fig.~\ref{fig:real_galaxies} shows images of the dust masks generated by our neural network for the $f_{mask}$
parameter value equal to 0.3. The next three rows show best model images for the simple decomposition (without dust
correction) first, then for the decomposition with a dust mask, and finally the decomposition with a dust model. It is evident that while the
two former models can, in general, reproduce the overall shape of the galaxy, they miss an absorption lane, while
the later one demonstrate the presence of a dust lane that is a part of the model. We note that the galaxy PGC\,27896
(left column) is not oriented perfectly edge on, but is slightly inclined which results in a shift of the dust lane below the
visible disc centre and in an asymmetry of the obscured bulge region. Both of these features are reproduced by our model that
has converged to an inclination value of 89 degrees.

The last three rows in Fig.~\ref{fig:real_galaxies} contain residual ``image -- model'' maps for the three decomposition methods. The
residuals for the decomposition without dust correction looks as expected, because
the model converges to some averaged disc+dust solution, in the galaxy plane the models are too bright, resulting in the residual maps
showing over-subtracted negative regions. Above and below the galactic plane, where the dust absorption is weaker, the
model is too faint, which results in bright regions in the residuals. The residual maps for
decompositions with a dust mask are different. Because the dust lanes of these galaxies are masked out, this does not lead to
attenuated disc models, and a brighter disc appears instead. As a result, the residual maps show overly subtracted regions
where the dust attenuation is high, but there are no bright regions above and below as in the previous case. This means that the
disc model better fit low absorption regions of images.

The residuals of the decomposition with our new {\small IMFIT} model are shown in the bottom row of Fig.~\ref{fig:real_galaxies}. They
demonstrate a better agreement between the observed images and the corresponding models. While there are still some deviations from the zero
level, they are considerably smaller. The dust lane is not overly subtracted and there are no bright regions
around the disc plane. Although there are still slightly  over-subtracted regions in the outer regions of the disc, they
are attributed to the fact that both galaxies have Freeman Type II discs \citep{Freeman1970}, whereas in our
models the disc is a pure exponential (Type I). In addition, \cite{2022MNRAS.512.1371M} catalogued these galaxies as having a B/PS bulge in the central part which is clearly visible in our residual images.

Numerical values for the decomposition results are listed in Table~\ref{tab:decomposition_real}, where for both tested
galaxies we present the fit structural parameters for a simple bulge+plus disc model (marked as ``not corrected'' in the
table), for a decomposition with a dust mask (``dust mask''), and for our model with a dust component (``dust
component''). One can see that for real galaxies we observe the same systematic shifts between the uncorrected and
corrected values (see Figs.~\ref{fig:bulge_no_corrections} and~\ref{fig:disc_no_corr}). After correction, the discs become
brighter, thinner, and have shorter radial scales, whereas the bulges become brighter (with higher bulge-to-total luminosity values) and
with larger S\'{e}rsic index values. We also mention that the results of the decomposition with a dust model have higher
discrepancies with uncorrected parameters than the results of the decomposition with a dust mask. The latter appears to be
somewhere in between the uncorrected parameters and the results of the decomposition with a dust model. This also aligns with the results of our numerical tests which demonstrate that the decomposition with dust masks of all kinds still
has higher errors and systematical shifts when compared with the dust model decomposition. This is especially clear for the
S\'{e}rsic parameters, which are almost the same as the uncorrected values.

\begin{table*}
  \centering
  \begin{tabular}[h]{|llccccccc|}
    \hline
                              &                        & $\mu_{0, \mathrm{disc}} $& $h_{\mathrm{disc}}$  & $z_{\mathrm{disc}}$ & $\mu_e$ & $r_e$  & $n$  & $B/T$\\
                              &                        & $mag/\square ''$     & ''                &   ''             & $mag/\square ''$ & ''   &\\
    \hline
    \multirow{2}{*}{PGC27896} &    Not corrected       & 20.00               &   9.8              &   2.2            &   20.8   & 2.19  &  0.4 & 0.12 \\
                              &    Dust mask           & 19.63               &   9.1              &   1.7           &   20.4   & 2.2   &  0.3 & 0.14 \\
                              &    Dust component      & 19.24               &   9.1              &   1.2           &   21.0   & 3.17  &  3.7 & 0.27 \\
    \hline
    \multirow{2}{*}{PGC2441449} &    Not corrected      & 20.26               &   8.9               &   2.1           & 21.33   & 3.3   &   0.6 &  0.19\\
                                &    Dust mask          & 19.54               &   7.5               &   1.5           & 21.20   & 3.3   &   0.7 & 0.20 \\
                                &    Dust component     & 19.21               &   7.0               &   1.1           & 20.42   & 3.2   &   3.6 &  0.39\\
  \end{tabular}
  \caption{Structural parameters of the two real galaxies recovered using a simple decomposition and the two methods of dust correction.}
  \label{tab:decomposition_real}
\end{table*}

\section{Conclusions}
\label{sec:conclusions}
In this article we ran a number of numerical simulations in order to determine how the presence of dust impacts the
measured values for the parameters of edge-on galaxies. To achieve this, we created a set of artificial galaxy images with
various parameters and applied a standard decomposition method to them to be able to compare the input and output values of the structural parameters.
We also tested three different techniques for how this impact can be minimized, two of which are based on masking dust
attenuated regions of galactic images, and the third involves an analytical model that includes dust absorption. Our main conclusions can be summarized as follows.

We confirm the findings of previous authors (\citealt{Gadotti2010, Pastrav2013a, Pastrav2013b}) who utilized two-dimensional decomposition to infer the general trends of how the bulge and disc parameters are altered due to the varied dust absorption. Using three-dimensional decomposition that accounts for the vertical structure of the stellar disc, we show that these trends hold true for perfectly edge-on galaxies, for which the disc thickness can not be neglected. For bulges, the measured values of the effective radius tend to be larger than the true (intrinsic) values, whereas the effective surface brightness, S\'ersic index,
and bulge-to-total luminosity ratio tend to be lower. In other words,  bulges in dusty edge-on galaxies appear fainter and less concentrated than they really are. For discs, the measured values of the central surface brightness tend to be lower than the true values, while the radial and vertical scales tend to be
larger. Therefore, discs also appear fainter and less concentrated than they are in reality. For both bulges and discs, the absolute values of the parameters' shifts depend on the properties of these components and on the dust content in the galaxy.

Masking out the regions most affected by dust in a galactic image allows one to considerably reduce the dust's
influence and obtain better estimates of the galactic parameters. Comparing different masking techniques showed that a
dust mask which is more extended in the center of the galaxy and is narrower in the outer region is better than a dust mask
which has a constant width. A neural network can be trained to effectively generate such masks for images of real
galaxies.

An analytical model that includes an absorbing dust component in a form of a 3D exponential disc can be used to
perform decomposition of a galaxy with a prominent dust lane and infer the galaxy structural parameters corrected for dust. Even if this model does not include light scattering, for simplification, it performs better than any
masking techniques that we tested.

The results of applying the proposed methods to a couple of real galaxies whilst taking the dust component into account
are in agreement with numerical experiments and demonstrate the validity of our approach.

We plan to continue our research of the dust impact on the decomposition of galaxies, and consider other wavelengths (such as ultraviolet and infrared ranges) as well as other, more complicated galaxy models. We also plan to continue the development of different algorithms to correct the decomposition results for the presence of dust 
in galaxies that are viewed in an orientation close to edge-on.

\section*{Acknowledgements}
We acknowledge financial support from the Russian Science Foundation (grant no. 20-72-10052).

%%%%%%%%%%%%%%%%%%%%%%%%%%%%%%%%%%%%%%%%%%%%%%%%%%

\section*{Data availability}
The data underlying this article will be shared on reasonable request to the corresponding author.



\bibliographystyle{mnras}
\bibliography{art}

%%%%%%%%%%%%%%%%%%%%%%%%%%%%%%%%%%%%%%%%%%%%%%%%%%

%%%%%%%%%%%%%%%%%%%%%%%%%%%%%%%%%%%%%%%%%%%%%%%%%%

%%%%%%%%%%%%%%%%% APPENDICES %%%%%%%%%%%%%%%%%%%%%

\appendix
\section{Neural networks for relative mask generation}
\label{sec:appendix_neural}
To train a neural network, one needs a dataset that can be split into  training, validating, and testing subdatasets.
To create such a dataset, we utilized the procedure based on the {\small SKIRT} package (see Sec.~\ref{sec:synthimag}). Each sample consists of four images of the same size ($256 \times 256$ pixels):
three model images in the $g$, $r$, and $i$ passbands of the SDSS system that were used as a neural network input, and a ground truth image of a relative dust mask. To make our model images more realistic, we overlapped them with random fields from the SDSS footprint which added background objects and noise. The parameters of modeled galaxies were randomly chosen with a uniform distribution in the following ranges:
\begin{itemize}
\item bulge-to-total luminosity ratio: 0.05 -- 0.35
\item bulge effective radius: 500 -- 1200 pc
\item bulge S\'ersic index: 0.5 -- 5
\item bulge ellipticity: 0.0 -- 0.5
\item stellar disc exponential scale: 2500 -- 5000 pc
\item stellar disc vertical scale: 0.1 -- 0.3 of a stellar radial scale
\item dust disc radial scale: 0.5 -- 1.5 of a stellar disc radial scale
\item dust disc vertical scale: 0.05 -- 0.3 of a stellar disc vertical scale
\item dust $\tau$ value: 0.1 -- 5.0.
\end{itemize}

We also allowed models to have inclinations in a range of $ \left[ 86: 90 \right] $ degrees to simulate galaxies that aren't viewed perfectly edge-on.

Since different relative masks (with different $f_{\mathrm{mask}}$ values) can lead to different results for the decomposition (as we demonstrate by our experiments, see below), we used a set of $f_{\mathrm{mask}}$ values equal to 0.1, 0.3, 0.4, and 0.5 generate datasets to train several independent neural networks that predict masks with such parameters. In total, we created 10000 samples in each dataset to train our networks.

 Each dataset was preprocessed in
the same manner. Here we briefly provide the main data preprocessing steps.
\begin{enumerate}
\item We split the dataset into training, validating and testing  as separate subdatasets of $8000 / 1000 / 1000$ images
  respectively.
\item Next, we calculated the common $99.9$th percentile values for each optical band ($g, r, i$) separately for the
  union of the training and validating subdatasets ($9000$ images). Then we applied corresponding clipping. This allowed us to
  moderate the problem associated with the brightest pixels which reduces the image contrast and decreases the training efficiency.
\item Then, we applied a natural logarithm transformation followed by min-max normalization to $\left[-1: 1\right]$ range.
\end{enumerate}

To train our neural network, we employed a sparse categorical cross-entropy loss function derived from the  output tensor with logit function. To
optimise the loss function, we chose the Adam optimization method. The batches consist of $32$ images. We trained all neural networks
for $60$ epochs on a single NVIDIA GeForce RTX 3060 GPU. The lowest validation loss is reached in $30$-$40$ epochs.

During the training experiments, we varied 2 parameters that influence the fitting process and the final model
performance: the training strategy (\guillemotleft training from scratch\guillemotright, \guillemotleft transfer
learning\guillemotright, or \guillemotleft fine-tuning\guillemotright) and the learning rate $r$.  In the \guillemotleft
transfer learning\guillemotright\,strategy, we took a pre-trained ImageNet dataset \citep{deng2009imagenet} encoder and
froze it before starting the training process.  In the \guillemotleft fine-tuning\guillemotright\,strategy, we also used a
pre-trained encoder but did not freeze it.

\label{sec:appendix}
\begin{table} 
	\caption{Results gathered from the conducted training experiments. This table lists the relative mask attenuation, training strategy (\guillemotleft training from scratch\guillemotright, \guillemotleft transfer learning\guillemotright, or \guillemotleft fine-tuning\guillemotright), learning rate, IoU, precision, and recall for all test galaxies for the relative mask class. The highest value for each metric is highlighted by the bold font for each relative mask.} 
	\centering
    \begin{tabular}{l l l c c c}
        \hline \hline
        $f_{\mathrm{mask}}$  & Training strategy & $r$ & IoU & precision & recall \\
\hline
$0.1$   &training from scratch    &$0.001$ &$0.823$ &$0.867$ &$\mathbf{0.942}$  \\
$0.1$   &training from scratch    &$0.0005$&$0.834$ &$0.92$  &$0.9$    \\
$0.1$   &training from scratch    &$0.00025$   &$0.826$ &$\mathbf{0.936}$ &$0.875$  \\
$0.1$   &training from scratch    &$0.0001$&$0.822$ &$0.923$ &$0.882$  \\
$0.1$   &transfer learning        &$0.001$ &$0.833$ &$0.929$ &$0.89$   \\
$0.1$   &transfer learning        &$0.0005$&$0.834$ &$0.913$ &$0.906$  \\
$0.1$   &transfer learning        &$0.00025$   &$0.834$ &$0.913$ &$0.906$  \\
$0.1$   &transfer learning        &$0.0001$&$0.824$ &$0.922$ &$0.885$  \\
$0.1$   &fine-tuning &$0.001$ &$0.837$ &$0.902$ &$0.921$  \\
$0.1$   &fine-tuning &$0.0005$&$\mathbf{0.838}$ &$0.915$ &$0.909$  \\
$0.1$   &fine-tuning &$0.00025$   &$0.836$ &$0.924$ &$0.897$  \\
$0.1$   &fine-tuning &$0.0001$&$0.83$  &$0.916$ &$0.899$  \\
\hline
$0.3$   &training from scratch    &$0.001$ &$0.821$ &$0.857$ &$\mathbf{0.951}$  \\
$0.3$   &training from scratch    &$0.0005$&$0.839$ &$\mathbf{0.94}$  &$0.886$  \\
$0.3$   &training from scratch    &$0.00025$   &$0.844$ &$0.916$ &$0.915$  \\
$0.3$   &training from scratch    &$0.0001$&$0.836$ &$0.914$ &$0.907$  \\
$0.3$   &transfer learning        &$0.001$ &$0.845$ &$0.907$ &$0.925$  \\
$0.3$   &transfer learning        &$0.0005$&$0.844$ &$0.92$  &$0.911$  \\
$0.3$   &transfer learning        &$0.00025$   &$0.843$ &$0.921$ &$0.909$  \\
$0.3$   &transfer learning        &$0.0001$&$0.837$ &$0.922$ &$0.901$  \\
$0.3$   &fine-tuning &$0.001$ &$\mathbf{0.847}$ &$0.925$ &$0.909$  \\
$0.3$   &fine-tuning &$0.0005$&$0.84$  &$0.928$ &$0.898$  \\
$0.3$   &fine-tuning &$0.00025$   &$\mathbf{0.847}$ &$0.919$ &$0.915$  \\
$0.3$   &fine-tuning &$0.0001$&$0.834$ &$0.932$ &$0.887$  \\
\hline
$0.4$   &training from scratch    &$0.001$ &$\mathbf{0.847}$ &$0.899$ &$\mathbf{0.935}$  \\
$0.4$   &training from scratch    &$0.0005$&$0.843$ &$0.931$ &$0.899$  \\
$0.4$   &training from scratch    &$0.00025$   &$0.846$ &$0.901$ &$0.933$  \\
$0.4$   &training from scratch    &$0.0001$&$0.828$ &$0.909$ &$0.902$  \\
$0.4$   &transfer learning        &$0.001$ &$0.84$  &$\mathbf{0.936}$ &$0.891$  \\
$0.4$   &transfer learning        &$0.0005$&$0.842$ &$0.926$ &$0.902$  \\
$0.4$   &transfer learning        &$0.00025$   &$0.84$  &$0.907$ &$0.92$   \\
$0.4$   &transfer learning        &$0.0001$&$0.833$ &$0.925$ &$0.894$  \\
$0.4$   &fine-tuning &$0.001$ &$0.836$ &$0.899$ &$0.922$  \\
$0.4$   &fine-tuning &$0.0005$&$0.843$ &$0.93$  &$0.901$  \\
$0.4$   &fine-tuning &$0.00025$   &$0.843$ &$0.933$ &$0.897$  \\
$0.4$   &fine-tuning &$0.0001$&$0.844$ &$0.929$ &$0.902$  \\
\hline
$0.5$   &training from scratch    &$0.001$ &$0.831$ &$0.898$ &$0.917$  \\
$0.5$   &training from scratch    &$0.0005$&$0.841$ &$0.928$ &$0.899$  \\
$0.5$   &training from scratch    &$0.00025$   &$0.839$ &$0.898$ &$\mathbf{0.928}$  \\
$0.5$   &training from scratch    &$0.0001$&$0.829$ &$0.919$ &$0.894$  \\
$0.5$   &transfer learning        &$0.001$ &$0.843$ &$0.919$ &$0.911$  \\
$0.5$   &transfer learning        &$0.0005$&$0.84$  &$0.918$ &$0.908$  \\
$0.5$   &transfer learning        &$0.00025$   &$0.836$ &$0.923$ &$0.898$  \\
$0.5$   &transfer learning        &$0.0001$&$0.829$ &$0.925$ &$0.888$  \\
$0.5$   &fine-tuning &$0.001$ &$0.838$ &$0.915$ &$0.909$  \\
$0.5$   &fine-tuning &$0.0005$&$0.838$ &$0.93$  &$0.894$  \\
$0.5$   &fine-tuning &$0.00025$   &$\mathbf{0.844}$ &$0.91$  &$0.92$   \\
$0.5$   &fine-tuning &$0.0001$&$0.831$ &$\mathbf{0.935}$ &$0.881$  \\

    \end{tabular} 
    \label{tab:nn_metrics}
\end{table}   
%\comm{

%}
%%%%%%%%%%%%%%%%%%%%%%%%%%%%%%%%%%%%%%%%%%%%%%%%%%

% Don't change these lines
\bsp	% typesetting comment
\label{lastpage}
\end{document}